\documentclass{emulateapj}

\submitted{Accepted for publication in ApJ}

\def\deg{\ifmmode^\circ\else$^\circ$\fi}

\def\gs{{_>\atop^{\sim}}}

\def\lsun{L$_{\odot}$}

\def\arcs{\ifmmode {''}\else $''$\fi}
\def\arcm{\ifmmode {'}\else $'$\fi}
\def\parcs{\sa=.07em \sb=.03em
     \ifmmode $\rlap{.}$^{\scriptscriptstyle\prime\kern -\sb\prime}$\kern -\sa$
     \else \rlap{.}$^{\scriptscriptstyle\prime\kern -\sb\prime}$\kern -\sa\fi}
\def\parcm{\sa=.08em \sb=.03em
     \ifmmode $\rlap{.}\kern\sa$^{\scriptscriptstyle\prime}$\kern-\sb$
     \else \rlap{.}\kern\sa$^{\scriptscriptstyle\prime}$\kern-\sb\fi}

\def\spose#1{\hbox to 0pt{#1\hss}}
\def\simlt{\mathrel{\spose{\lower 3pt\hbox{$\mathchar"218$}}
     \raise 2.0pt\hbox{$\mathchar"13C$}}}
\def\simgt{\mathrel{\spose{\lower 3pt\hbox{$\mathchar"218$}}
     \raise 2.0pt\hbox{$\mathchar"13E$}}}
\def\lsim{\rlap{$<$}{\lower 1.0ex\hbox{$\sim$}}}
\def\gsim{\rlap{$>$}{\lower 1.0ex\hbox{$\sim$}}}

\usepackage[usenames]{color}

\begin{document}

\title{PAH Emission from Ultraluminous Infrared Galaxies\altaffilmark{1}}

\author{V. Desai\altaffilmark{2},
L. Armus\altaffilmark{3},
H.W.W. Spoon\altaffilmark{4},
V. Charmandaris\altaffilmark{5,6},
J. Bernard-Salas\altaffilmark{4},
B.R. Brandl\altaffilmark{7},
D. Farrah\altaffilmark{4},
B.T. Soifer\altaffilmark{2,3},
H.I. Teplitz\altaffilmark{3},
P.M. Ogle\altaffilmark{3},
D. Devost\altaffilmark{4},
S.J.U. Higdon\altaffilmark{8},
J.A. Marshall\altaffilmark{4},
J.R. Houck\altaffilmark{4}
}

\altaffiltext{1}{Based on observations obtained with the \textit{Spitzer Space
Telescope}, which is operated by the Jet Propulsion Laboratory,
California Institute of Technology, under NASA contract 1407}

\altaffiltext{2}{Division of Physics, Math \& Astronomy, California
Institute of Technology, Pasadena, CA 91125}

\altaffiltext{3}{Spitzer Science Center, MS 220-6, Caltech, Pasadena, CA 91125}

\altaffiltext{4}{Cornell University, Ithaca, NY 14853}

\altaffiltext{5}{University of Crete, Department of Physics, P.O. Box
2208 GR-71003, Heraklion, Greece}

\altaffiltext{6}{IESL/Foundation for Research and Technology - Hellas,
GR-71110, Heraklion, Greece and Chercheur Associ\'e, Observatoire de
Paris, F-75014, Paris, France}

\altaffiltext{7}{Leiden University, P.O. Box 9513, 2300 RA Leiden, The
Netherlands}

\altaffiltext{8}{Georgia Southern University, Statesboro, GA 30460}

\altaffiltext{9}{The IRS was a collaborative venture between Cornell
University and Ball Aerospace Corporation funded by NASA through the
Jet Propulsion Laboratory and the Ames Research Center}

\begin{abstract}

We explore the relationships between the Polycyclic Aromatic
Hydrocarbon (PAH) feature strengths, mid-infrared continuum
luminosities, far-infrared spectral slopes, optical spectroscopic
classifications, and silicate optical depths within a sample of 107
ULIRGs observed with the Infrared Spectrograph on the {\it Spitzer
Space Telescope}.  The detected 6.2 $\micron$ PAH equivalent widths
(EQWs) in the sample span more than two orders of magnitude
($\sim$0.006--0.8 $\micron$), and ULIRGs with HII-like optical spectra
or steep far-infrared spectral slopes ($S_{25} / S_{60} < 0.2$)
typically have 6.2 $\micron$ PAH EQWs that are half that of
lower-luminosity starbursts.  A significant fraction ($\sim$40--60\%)
of HII-like, LINER-like, and cold ULIRGs have very weak PAH EQWs.
Many of these ULIRGs also have large ($\tau_{9.7} > 2.3$) silicate
optical depths.  The far-infrared spectral slope is strongly
correlated with PAH EQW, but not with silicate optical depth.  In
addition, the PAH EQW decreases with increasing rest-frame 24
$\micron$ luminosity.  We argue that this trend results primarily from
dilution of the PAH EQW by continuum emission from dust heated by a
compact central source, probably an AGN.  High luminosity,
high-redshift sources studied with {\it Spitzer} appear to have a much
larger range in PAH EQW than seen in local ULIRGs, which is consistent
with extremely luminous starburst systems being absent at low
redshift, but present at early epochs.

\end{abstract}


\section{Introduction}
\label{sec:Introduction}

Ultraluminous Infrared Galaxies (ULIRGs) have bolometric luminosities
comparable to quasars (${\rm L}_{\rm bol} \gs 10^{12}$ \lsun), but
emit nearly all of this energy at mid- and far-infrared wavelengths.
Although ULIRGs are rare in the local Universe and comprise only a few
percent of infrared-bright galaxies \citep{Soifer87,Sanders03}, they
account for a rapidly increasing fraction of all star-formation
activity at high-redshift, and they may dominate the far-infrared
background at $z > 2$
\citep{Franceschini01,LeFloch05,PerezGonzalez05}.  It has been
suggested that they play a role in the formation of both quasars and
elliptical galaxies \citep{Kormendy92,SandersMirabel96,Scott02}.

To understand the roles that ULIRGs play in the formation of massive
galaxies, the global star formation history of the universe, and the
creation of the cosmic infrared background, it is essential to
determine the extent to which Active Galactic Nuclei (AGN) and star
formation contribute to their enormous luminosities.  Several studies
have used optical
\citep{deGrijp85,Osterbrock85,Armus87,Armus89,Veilleux95,Kim98,Veilleux99}
and near-infrared
\citep{Goldader95,Veilleux97,Veilleux99b,Murphy99,Murphy01,Burston01,Davies03,Dannerbauer05}
spectroscopy to determine the main source of ionization of the
line-emitting gas in local ULIRGs.  These studies show that while the
majority of ULIRGs have optical and near-infrared spectra similar to
those of starburst galaxies, the fraction of ULIRGs displaying the
spectroscopic signatures of an AGN, either broad permitted lines or
high-ionization narrow-lines, increases with infrared luminosity.  In
addition, evidence for AGN activity is more commonly found among
ULIRGs with flatter far-infrared spectral slopes (those classified as
``warm" with $ S_{25} / S_{60} \ge 0.2$ as measured with IRAS), than
among ULIRGs with steeper far-infrared spectral slopes (those
classified as ``cold" with $S_{25} / S_{60} < 0.2$).

A principle uncertainty of these optical and near-infrared studies is
the extent to which the physical conditions of the visible gas are
linked to those in the deeply embedded nucleus, where the bulk of the
energy is generated \citep{Soifer00}.  Emission at longer wavelengths
can penetrate more heavily obscured regions close to the nuclear power
source.  Therefore, diagnostics derived from mid-infrared spectroscopy
are potentially a better probe of the dominant energy source in the
nuclear regions. The mid-infrared spectra of ULIRGs are comprised of
continuum emission from heated dust grains; broad emission features
associated with Polycyclic Aromatic Hydrocarbons (PAHs), the most
prominent of which appear at 6.2, 7.7, 8.6, 11.3, and 12.7 $\micron$;
absorption by amorphous silicates centered at 9.7 and 18 $\micron$;
and atomic fine-structure lines of Ne, O, Si, and S covering a large
range in ionization potential.  Diagnostic diagrams using combinations
of fine-structure line ratios, mid-infrared spectral slope, and PAH
feature strengths have been used to classify bright ULIRGs, based on
the expectations that ULIRGs with central AGN produce high ionization
lines, have flatter spectral slopes, and display lower PAH equivalent
widths than those without
\citep[e.g.][]{Genzel98,Lutz98,Rigopoulou99,Laurent00,Tran01,Sturm02,Armus04,Armus06,Armus07,Farrah07}.
For sources at high-redshift ($z \ge 1$), however, it is generally not
possible to place useful limits on the high-ionization fine-structure
lines.  The PAH emission, together with the infrared spectral slope,
are often the only mid-infrared diagnostics available for studying
high-redshift ULIRGs.

The first attempts to classify the energy generation mechanism in
ULIRGs using mid-infrared PAH features made use of data from the
\textit{Infrared Space Observatory}
\citep{Lutz98,Lutz99,Rigopoulou99,Tran01}.  For these studies, ULIRGs
were classified based on the strengths of their 7.7 $\micron$ PAH
line-to-continuum ratios (L/C, defined as the ratio of the peak height
of the 7.7 $\micron$ PAH feature to the level of the underlying 7.7
$\micron$ continuum).  The PAH-based classifications showed that most
($\sim$80\%) ULIRGs have strong PAH emission indicative of star
formation as the dominant power source.  However, the highest
luminosity sources tend to have weak (or undetected) PAH emission,
interpreted as a sign of increased AGN activity.  Recently,
\citet{Imanishi07} analyzed the mid-infrared spectra of 48 nearby
ULIRGs observed with the Infrared Spectrograph \citep[IRS;][]{Houck04}
on board the \textit{Spitzer Space Telescope}.  These authors suggest
that 30-50\% of these ULIRGs, all of which have HII-like or LINER-like
optical classifications, harbor buried AGN.

In this paper we present the mid-infrared spectroscopic properties of
a large sample of 107 ULIRGs that comprise the IRS GTO team ULIRG
survey.  The sample includes a large fraction ($\sim$40\%) of warm
sources, which tend to have higher infrared luminosities than cold
ULIRGs.  The warm ULIRGs are a critical population to study, since it
has been suggested that they mark the transition between
staburst-dominated ULIRGs and QSOs \cite[e.g.~][]{Sanders88b}.  Because
of their high luminosities, they provide a much-needed baseline over
which to extrapolate to the extremely high luminosities (L$_{\rm IR} >
10^{13}$ L$_{\odot}$) seen at higher redshift.

The sensitivity of the IRS offers two distinct advantages over
previous studies with ISO.  First, we can measure the PAH strengths
with respect to the dust continuum (the equivalent width, or hereafter
EQW) over nearly two orders of magnitude, allowing us to quantify the
PAH emission in many ULIRGs that previously only had upper
limits. This is critical for fitting trends with luminosity and
spectral slope, as many ULIRGs have much weaker PAH emission features
than are typically found in low-luminosity starburst galaxies (see
\S{\ref{sec:medianspectra}}).  Second, the Short-Low and Long-Low IRS
modules (5--38 $\micron$) offer greatly increased wavelength coverage
compared to ISOPHOT-S, allowing us to accurately fit the silicate
absorption (at 9.7 and 18 $\micron$) and multiple PAH features in all
the ULIRGs.  Here, we have chosen to fit the 9.7 $\micron$ silicate
absorption feature since it is the strongest and the 6.2 and 11.3
$\micron$ PAH features because they are both well-isolated and easily
measured.  Previously the 7.7 $\micron$ feature was often used, which
is not ideal because absorption at both shorter and longer wavelengths
can mimic a peak at 7.7 $\micron$ \citep{Spoon02} and its integrated
flux can be difficult to measure due to blending with the adjacent 8.6
$\micron$ feature.  By correlating the PAH emission with the silicate
absorption, it is possible to gain insight into the distribution of
the heating source(s) and the grains.  In the following analysis, we
use ${\rm H_0}=70$ km s$^{-1}$ Mpc$^{-1}$, $\Omega_{\rm m}=0.3$, and
$\Lambda=0.70$.

\section{Observations \& Data Reduction}
\label{sec:Observations}
 
The 107 ULIRGs presented in this paper were chosen primarily from the
IRAS 1-Jy survey \citep{Kim98}, the IRAS 2-Jy survey
\citep{Strauss92}, and the FIRST/IRAS radio-far-infrared sample of
\citet{Stanford00}.  The ULIRGs span a redshift range of $0.018 < z <
0.93$, have IRAS 60 $\micron$ flux densities between $0.14 < {\rm
S}_{60} / {\rm Jy} < 103$, and have integrated 40--500 $\micron$
luminosities in the range $11.7 < \log({\rm L}_{\rm FIR} / {\rm
L}_{\odot}) < 13.13$.  The sample consists of 41 warm ULIRGs and 66
cold ULIRGs, based upon their rest-frame far-infrared flux densities
at 25 and 60 $\micron$.  Many of our highest luminosity sources were
taken from the \citet{Stanford00} FIRST/IRAS survey.  While
radio-bright AGN may preferentially populate the highest luminosities
of the Stanford et al.~survey, the selected ULIRGs have radio to
far-infrared flux density ratios consistent with infrared-selected
starburst galaxies of lower luminosity.  The sample is specifically
designed to include a larger fraction of warm and infrared luminous
sources than would be found in a pure flux- or volume-limited sample,
for the express purpose of allowing a careful study of the
mid-infrared spectral properties as functions of infrared color and
luminosity.

All ULIRGs were observed in Staring Mode with both sub-slits (orders)
of each of the Short-Low (SL) and Long-Low (LL) modules of the IRS.
Each target was acquired by performing a high accuracy IRS peak-up on
the target itself, or by peaking up on a nearby 2MASS star and
offsetting to the target.  Each galaxy was observed at two nod
positions within each of the IRS sub-slits (SL1, SL2, LL1, LL2). The
resulting spectra have a spectral resolution of R $\sim 80$ over the
5--38 $\micron$ wavelength range.

All spectra were reduced using the S14 IRS pipeline at the Spitzer
Science Center.  This reduction includes ramp fitting, dark sky
subtraction, droop correction, linearity correction, and wavelength
and flux calibration.  The flux calibration sources were HD173511 for
the SL and LL2 modules and KsiDra for the LL1 module. For a given
module, order, and nod position, the background in the two-dimensional
spectrum was removed by subtracting the combined two-dimensional data
taken with the same module, but adjacent order.  One-dimensional
spectra were extracted from the background-subtracted two-dimensional
spectra using the SMART data reduction package \citep{Higdon04}.  The
adopted extraction apertures are four pixels at the blue end of each
order and expand linearly with wavelength.

The 6.2 and 11.3 $\micron$ PAH features were measured by integrating
the flux above a spline-interpolated continuum.  For those sources
with strong 6 $\micron$ water ice absorption, the apparent 6.2
$\micron$ EQW has been corrected by using an inferred 6.2 $\micron$
continuum defined by a spline interpolation between 5--26 $\micron$,
using pivot points at 5.2, 5.6, 7.8 (in the case of very weak PAH
emission), 14, and 26 $\micron$.  This correction results in a lower
6.2 $\micron$ PAH EQW, and is correct under the assumption that the
PAH emission is unaffected by the bulk of the ice absorption.  A full
description of the ice-fitting procedure is given by \cite{Spoon07},
who first present the measured 6.2 $\micron$ EQWs for many of these
ULIRGs (see \S 3.2 and Figure 2 for a description of the correction
and Figure 1 for an indication of the size of the correction).  The
PAH emission at 11.3 $\micron$ is affected by the broad 9.7 $\micron$
silicate absorption feature, which can be deep in ULIRGs.  However,
lacking a strong understanding of how the obscuring dust is
distributed, we have not attempted to correct for it.  Of the 107
ULIRGs analyzed in this paper, the 6.2 $\micron$ PAH feature is
detected for all but 11 sources, and the 11.3 $\micron$ PAH feature is
detected for all but 12 sources.

In the following, we investigate the variation of the PAH features
with rest-frame continuum fluxes at 5.5, 24, 25, and 60 $\micron$.
The 5.5 $\micron$ flux is taken to be the average continuum betweeen
5.3 and 5.8 $\micron$.  The 24 $\micron$ flux was calculated by
convolving the IRS spectrum with the MIPS 24 $\micron$ bandpass.  In
the 10 cases where the IRS spectrum did not extend over the full
wavelength range of the wide IRS bandpass due to trimming of pipeline
artifacts (e.g.~fringes), it was extended by linearly interpolating in
log-log space between the IRS spectrum and the 60 $\micron$ IRAS flux
density \citep{Moshir90}.  No color-correction was made to place these
filter fluxes on the MIPS absolute scale, but these corrections are
small ($<$5\%) based on the spectral shape of the ULIRGs within the 24
$\micron$ bandpass (see the MIPS Data Handbook).  The 25 $\micron$
rest-frame flux was estimated by taking the median of the trimmed and
(when needed) interpolated IRS spectrum between 24.7 and 25.3
$\micron$.  The rest-frame 60 $\micron$ luminosity was estimated by
linearly interpolating in log-log space between the observed IRAS 60
and 100 $\micron$ flux densities.  For 12 ULIRGs, this interpolation
likely underestimates the 60 $\micron$ luminosity, either because the
object is at a redshift high enough ($z > 0.65$) that the observed 100
$\micron$ IRAS observation corresponds to a rest-frame wavelength less
than 60 $\micron$, or because the 100 $\micron$ observation resulted
in only a limit, rather than a detection.

Classifications based on optical spectra are available from the
literature for 64 of the 107 ULIRGs in our sample.  Of these, 15 have
HII-like optical spectra, 22 are LINER-like, and 27 are Seyfert-like.

\section{Results}

\subsection{PAH strength and silicate absorption versus spectral slope and optical classification}
\label{sec:medianspectra}

The median IRS spectra for warm and cold ULIRGs are shown in the
left-hand panel of Figure \ref{fig:medianspectra_warmcold}.  Cold
ULIRGs have a median 6.2 $\micron$ PAH EQW that is a factor of six
higher than the median measured for warm ULIRGs (0.24 versus 0.04
$\micron$; see Table \ref{table:pahew_v_class}). The right-hand panel
of Figure \ref{fig:medianspectra_warmcold} shows the average starburst
spectrum from \citet{Brandl06} plotted over the median cold ULIRG
spectrum.  Cold ULIRGs have a median 6.2 $\micron$ PAH EQW which is
about half the median for lower-luminosity starbursts (see also
\citet{Imanishi07}).  The difference in the PAH EQWs in the median
warm and cold ULIRG spectra reflects the strong trend between PAH EQW
and mid-infrared spectral type shown in Figure \ref{fig:pah_v_slope}.
There is a huge range in measured PAH EQW among the sample ULIRGs.
The full range in PAH EQW for those ULIRGs with detected emission is
0.006--0.864 $\micron$ for the 6.2 $\micron$ feature, and 0.006--1.169
$\micron$ for the 11.3 $\micron$ feature.  The smallest upper limits
are 0.005 and 0.004 $\micron$ for the 6.2 and 11.3 $\micron$ features,
respectively.  While the median 6.2 $\micron$ PAH EQW among cold/HII
ULIRGs is about half as large as for local starburst galaxies, the
median 11.3 $\micron$ PAH EQW in cold/HII ULIRGs is 80\% of that
measured in starburst galaxies.  The cold ULIRGs and starburst
galaxies track each other over the limited range of $ 0.06 <S_{25} /
S_{60} < 0.2$ (see Figure \ref{fig:pah_v_slope}).

Figure \ref{fig:medianspectra_warmcold} also demonstrates that the
median cold source shows stronger silicate absorption at both 9.7 and
18 $\micron$ than the median warm source (1.36 versus 0.65 for
$\tau_{9.7}$, the optical depth at 9.7 $\micron$), and the median cold
ULIRG has stronger absorption than the average starburst (1.36 versus
0.22).  In Figure \ref{fig:tau_v_slope}, we plot the apparent silicate
optical depth as a function of far-infrared spectral slope.  While the
ULIRGs with $\tau_{9.7} > 1$ have steeper far-infrared spectra than
the sources with $\tau_{9.7} < 1$, we find no real correlation (see
also \citet{Imanishi07,Hao07}).  In fact, many sources with very high
silicate optical depths ($\tau_{9.7} > 2$) would also be classified as
warm ULIRGs.  All of the ULIRGs with Seyfert-1 optical spectra, and
many, but not all, of the ULIRGs with Seyfert-2 optical spectra have
$\tau_{9.7} < 1$ and $ S_{25} / S_{60} > 0.2$.

The median HII, LINER, and Seyfert ULIRG IRS spectra are shown in the
left-hand panel of Figure \ref{fig:medianspectra_optical}. The median
and mean values of the 6.2 and 11.3 $\micron$ PAH EQW for each of
these spectral classes is given in Table \ref{table:pahew_v_class}.
As expected, the HII-like ULIRGs have the largest median 6.2 $\micron$
PAH EQW (0.28 $\micron$) while the ULIRGs classified as Seyferts have
the lowest (0.04 $\micron$).  The median PAH EQW for infrared cold
ULIRGs is comparable to the median for those classified as HII-like,
while the median PAH EQW for infrared warm sources is similar to the
median for Seyfert-like ULIRGs.  There are a number of ULIRGs with
HII-like or LINER-like spectra that have unusually low 6.2 $\micron$
PAH EQW ($\lsim$0.15 $\micron$, or about 1/4 that found for pure
starburst galaxies).  Of the 22 LINER-like (15 HII-like) ULIRGs in the
sample, 12 (six) have 6.2 $\micron$ PAH EQW $\le 0.15$ $\micron$.
Similarly, of the 66 cold ULIRGs, 23 have a 6.2 $\micron$ PAH EQW $\le
0.15$ $\micron$.

The right-hand panel of Figure \ref{fig:medianspectra_optical} shows
the median HII and LINER ULIRG spectra overplotted with the average
starburst spectrum from \citet{Brandl06} and the average
infrared-bright LINER spectrum from \citet{Sturm06}.  Although sources
with HII-like optical spectra have the largest PAH EQWs among ULIRGs,
the median values are only about $50\%$ of those measured for pure
starburst galaxies, the same result as found for the cold ULIRGs.
HII-like ULIRGs also have stronger silicate absorption at 9.7 and 18
$\micron$ than seen in lower-luminosity starbursts (1.52 versus 0.22
for $\tau_{9.7}$).  Compared to infrared-bright LINERs, LINER-like
ULIRGs have $\sim 25$\% smaller PAH EQWs, and stronger silicate
absorption ($\tau_{9.7} = 2.11$ for the LINER-like ULIRGs while it is
only 0.84 for the IR-bright LINERs).

\subsection{PAH equivalent width versus continuum luminosity}
\label{sec:pahew_v_lum}

In Figure \ref{fig:pahew_v_lum24}, we plot the 6.2 and 11.3 $\micron$
PAH EQW versus 24 $\micron$ rest-frame luminosity.  This continuum
wavelength was chosen because it samples the peak emission of warm
($\sim$120 K) thermally-emitting dust, it can be measured much more
accurately than LIR in many cases, it is free from strong emission or
absorption features, and it allows direct comparison to a variety of
low-redshift samples observed with the MIPS 24 $\micron$ filter and to
$z \sim 2$ samples observed with the MIPS 70 $\micron$ filter.  For
reference, we estimate that $\log_{10}(\nu L_{\nu}(24 \micron) / {\rm
LIR}) = -1.06 \pm 0.47$ for cold ULIRGs and $-0.79 \pm 0.38$ for warm
ULIRGs.  These estimates were made by applying the
\citet{SandersMirabel96} prescription for converting IRAS photometry
into LIR to the nearest ($z < 0.2$) ULIRGs in our sample.  Figure
\ref{fig:pahew_v_lum24} shows a trend of decreasing 6.2 and 11.3
$\micron$ PAH EQW with increasing rest-frame 24 $\micron$ luminosity.
There is significant scatter in this relation, and ULIRGs with low PAH
EQW span a wide range in luminosity.  Nevertheless, the presence of a
correlation is verified by a highly significant Spearman correlation
coefficient, whether or not upper limits are included in the same way
as detections in the calculation.

We have divided the ULIRG sample into six luminosity bins chosen to
span the full range in 24 $\micron$ luminosity ($(0.03-8)$ $\times$
10$^{12}$ L$_{\odot}$), and to contain approximately equal numbers
(17-18) of ULIRGs.  The median 6.2 and 11.3 $\micron$ PAH EQWs drop by
an order of magnitude between the first and last luminosity bins (see
Table \ref{table:pahew_v_lum}).  The median ULIRG spectrum in each 24
$\micron$ luminosity class is shown in Figure \ref{fig:medianspectra}.
The median spectra were constructed using all sources in a given
luminosity bin, even those for which only an upper limit on the PAH
EQW could be measured.  Moving from lowest to highest 24 $\micron$
luminosity (top to bottom in Figure \ref{fig:medianspectra}), the PAH
features diminish and the spectrum flattens.  While the median
spectrum in the highest luminosity bin has a noticeably smaller
silicate absorption than the lower-luminosity median spectra, there is
a large range in the apparent silicate optical depth among the most
luminous sources.  This is consistent with the results of
\citet{Spoon07}, who find a dichotomy in silicate optical depths among
sources with small 6.2 $\micron$ PAH EQW (see also
\S{\ref{sec:tau_v_pahew}).

\subsection{Silicate absorption versus PAH equivalent width}
\label{sec:tau_v_pahew}

Figure \ref{fig:fork} shows the silicate optical depth versus 6.2
$\micron$ PAH EQW for the ULIRGs in our sample.  Such a plot was first
presented by \citet{Spoon07}, who discussed the presence of two
discrete tracks, or branches.  Figure \ref{fig:fork} shows that the
horizontal branch is populated almost exclusively by infrared warm
sources, while the diagonal branch is populated by about half of the
warm sources and virtually all of the cold sources.  Examining the
ULIRGs for which we have optical spectroscopic classifications, the
diagonal branch harbors about half of the Seyfert-like ULIRGs, and
nearly all of the HII-like and LINER-like ULIRGs.  The horizontal
branch is populated mainly by ULIRGs showing Seyfert-like spectra,
including all of the Type I sources.  ULIRGs with Seyfert Type 2
optical spectra are found on both branches, spanning (nearly) the full
range in silicate optical depth.  Most of the ULIRGs without optical
spectroscopic classifications lie on the diagonal branch.  As noted in
\S{\ref{sec:medianspectra}}, a significant fraction of the HII and
LINER-like ULIRGs have very low 6.2 $\micron$ PAH EQWs ($\le$0.15
$\micron$).  Of these, almost all of the LINERs (11/12) and half (3/6)
of the HII-like ULIRGs have $\tau_{9.7} \ge 2.5$, implying A$_{V} \ge
45$ mag.

\section{Discussion}
\label{sec:Discussion}

The median 6.2 $\micron$ PAH EQWs of cold (or HII-like) ULIRGs is
about 50\% of that seen in lower-luminosity starbursts.  Warm ULIRGs
have 6.2 $\micron$ PAH EQWs that are up to two orders of magnitude
smaller than those found in starbursts.  There are several possible
explanations for the extremely large range in PAH EQW and the overall
low PAH EQW in ULIRGs compared to starbursts.  Some ULIRGs may host a
central AGN, and the associated soft X-ray and ultraviolet radiation
may destroy PAH molecules \citep[e.g.][]{Aitken85,Voit92}, or
thermalize the emission.  The AGN may also heat dust to significant
temperatures ($>$400 K), resulting in excess continuum emission which
would lower the measured 6.2 $\micron$ PAH EQW, in efffect diluting
the starburst-driven PAH emission with respect to the AGN-powered, hot
dust continuum.  For either of these explanations, $55\pm13$\% of the
LINER-like ULIRGs and $40\pm15$\% of the HII-like ULIRGs in our sample
would harbor buried AGN.  In comparison, \citet{Lutz99} find the
percentage of LINER-like ULIRGs classified by ISO as AGN to be
$17\pm15$\% and the percentage of HII-like ULIRGS classified by ISO as
AGN to be $20\pm12$\%.  We are thus finding a much higher fraction of
LINERs and HII-like ULIRGs with weak PAH emission, although, given the
small sample sizes and the fact that the percentage is comparable to
the statistical errors in the ISO sample, our results are formally
consistent with those of \citet{Lutz99} at the 2$\sigma$ level.  Our
results are more comparable to those of \citet{Imanishi07}, who find
that 30--50\% of HII- and LINER-like ULIRGs contain dominant buried
AGN. Of the 12 LINER-like and 6 HII-like ULIRGs with low PAH EQWs, 9
were observed with the high-resolution modules of the IRS
\citep{Farrah07}.  Of these, none have detectable [NeV]14.3$\micron$,
[NeV]24.3$\micron$, or [OIV]24.9$\micron$ line emission.  Upper limits
on the [NeV]14.3$\micron$ to [NeII]12.8$\micron$ line flux ratio range from
0.02 to 0.3, corresponding to starburst-dominated ULIRGs with less
than $5-25$\% AGN contribution \citep{Armus07}.
However, it is important to note that many of these sources are deeply
buried, as evidenced by their large silicate optical depths (see
\S{\ref{sec:tau_v_pahew}).  As a result, the
[NeV]14.3$\micron$ emission may be highly extinguished.  In addition,
it is plausible that a completely buried starburst exists in these
sources, as in NGC 1377 \citep {Roussel06}.

The 6.2 $\micron$ PAH EQW in ULIRGs could also be (intrinsically)
smaller than that found in pure starburst galaxies if star formation
in ULIRGs is accompanied by larger quantities of hot dust emission
than is typically found in lower-luminosity starbursts.  The dust
would then have to ``see" more of the stellar radiation and compete
more effectively for UV photons, thus raising its average temperature,
something that is not unreasonable given the compact nature of the
infrared emission in ULIRGs. However, if this were the case, then both
the 6.2 and the 11.3 $\micron$ PAH EQW would be similarly depressed.
However, among cold ULIRGs, the median 11.3 $\micron$ PAH EQW is 80\%
of that found in starbursts, while the median 6.2 $\micron$ PAH EQW is
only 50\%.  A combination of dilution and extinction offers a more
likely explanation.  In this scenario, the 6.2 $\micron$ PAH EQW is
diluted by hot dust continuum produced by a central ionizing source
interior to the star-forming regions.  The dilution of the 11.3
$\micron$ PAH EQW is less significant because of its proximity to the
broad silicate absorption feature at 9.7 $\micron$.  The hot dust
continuum responsible for diluting the 6.2 $\micron$ PAH EQW is
heavily extincted at 11.3 $\micron$, and therefore does not readily
dilute the 11.3 $\micron$ PAH EQW.  This is seen in Figure
\ref{fig:fork}.  The left-hand panels show that the 6.2 $\micron$ PAH
EQWs of ULIRGs on the upper branch deviate strongly from the
starbursts.  In contrast, the right-hand panels indicate that ULIRGs
have nearly starburst-like 11.3 $\micron$ PAH EQWs because the strong
extinction at this wavelength overwhelms all but the strongest
contribution from hot dust.

Finally, extinction is not likely to play a large role in the reduced
6.2 $\micron$ PAH EQWs found in ULIRGs compared to starbursts.
Although ULIRGs on average display stronger silicate absorption
features than lower-luminosity starbursts, larger column densities of
dust would lead to depressed PAH EQWs only if the PAH emission were
more extinguished than the continuum.  Such a situation is unlikely,
since the continuum flux is emitted by hot dust that must lie close to
the ionization source.

In a further attempt to disentangle reduced PAH emission from
increased hot dust emission as the driving force behind the low 6.2
$\micron$ PAH EQWs in ULIRGs, we plot the 6.2 $\micron$ PAH luminosity
against the rest-frame 5.5 $\micron$ luminosity in Figure
\ref{fig:pahlum_v_lum5p5}.  Low-luminosity starbursts from
\citet{Brandl06} follow a correlation between these two quantities,
characterized by $\log_{10}$[L(6.2 $\micron$ PAH)] = -0.49 + 0.96
$\times$ $\log_{10}[\nu L_{\nu}(5.5 \micron)]$.  ULIRGs do not lie on
the extrapolation of this relation to higher luminosities. Instead,
ULIRGs appear offset to higher 5.5 $\micron$ luminosities for their
measured 6.2 $\micron$ PAH luminosities.  Not surprisingly, sources
with flatter far-infrared spectra and those with Seyfert-like optical
spectra are, as a group, the most displaced towards higher 5.5
$\micron$ luminosity.  However, the range of 6.2 $\micron$ PAH
luminosity within the different classes (both optical and infrared) is
the same.  It has been suggested that an evolutionary link exists
between ULIRGs and QSOs, such that cold ULIRGs evolve into warm ULIRGs
on their way to becoming QSOs \citep[e.g.][]{Sanders88}.  Figure
\ref{fig:pahlum_v_lum5p5} implies that the transition between cold and
warm ULIRGs involves primarily an increase in hot dust, rather than
the suppression of PAH emission. \citet{Lutz98} also suggested that
dilution of the EQW by hot dust was the dominant effect among ULIRGs.

Also plotted in the lower left-hand panel of Figure
\ref{fig:pahlum_v_lum5p5} are $\sim$80 PG QSOs at $z < 0.5$ (Ogle \&
Antonucci 2007, in preparation; Shi et al. 2007, submitted).  Most
have only upper limits for the 6.2 micron PAH EQW, but 15 have
detected values.  The 5.5 $\micron$ continuum
luminosities of ULIRGs and PG QSOs span the same range.  
The 15 PG QSOs with detected 6.2 $\micron$ PAH
features show no evidence for a correlation in Figure
\ref{fig:pahlum_v_lum5p5}.  This is in contrast with the results of
\citet{Schweitzer06}, who find a correlation between the 7.7 $\micron$
PAH luminosity and the 6 $\micron$ continuum luminosity among a sample
of 25 PG QSOs, including 15 with upper limits on the PAH luminosity
and 11 with detections.  Based on the assumptions that the 6 $\micron$
luminosity is due to hot dust heated by an AGN, while the PAH
luminosity is emitted by star-forming regions, they argue that their
observed correlation is evidence for a starburst-AGN connection among
PG QSOs.  However, we find no such correlation when considering the 6.2
$\micron$ PAH feature.

While the flatter spectra and lower 6.2 $\micron$ PAH EQWs of many of
the ULIRGs can readily be explained by an excess of hot dust emission,
the silicate optical depth, at least in some cases, seems inconsistent
with a simple model for the emission.  The 9.7 $\micron$ silicate
optical depth in ULIRGs varies from $-0.2 < \tau_{9.7} < 4$, and is on
average deeper than in starburst galaxies.  The radiative transfer
models of \citet{Levenson07} suggest that the geometry of the
obscuring material in ULIRGs determines both the depth of the silicate
feature and the infrared spectral slope.  According to these models,
the large apparent silicate optical depth ($\tau_{9.7} >$ 1--2)
observed in some ULIRGs requires an optically and geometrically thick,
smooth dust component, since a clumpy dust distribution can only
result in a shallow ($\tau_{9.7} < 1$) absorption feature.  All of the
Seyfert-like ULIRGs would then be viewed through this clumpy dust
distribution, exhibiting flat infrared spectra, weak PAH EQWs, and
shallow (or nonexistent) silicate absorption features.  If this simple
model is correct, a correlation between silicate optical depth and
spectral slope might be expected.  However, we find no such
correlation (see Figure \ref{fig:tau_v_slope}).  In fact, a
significant number of the sources with the most extreme absorption are
warm (see Figure \ref{fig:fork}).  These ULIRGs also tend to have low
6.2 $\micron$ PAH EQWs ($<0.15 \micron$).  A majority of the sources
with large silicate optical depth are classified optically as LINERs.

Among local ULIRGs, the sources with the highest rest-frame 24
$\micron$ luminosities have the lowest 6.2 $\micron$ PAH EQWs (see
Figure \ref{fig:pahew_v_lum24}).  The fit to this correlation is shown
as the black line in Figure \ref{fig:pahew_v_lum24_medians}, and is
given by $\log_{10}(6.2 \ \micron \ {\rm PAH \ EQW} [\micron]) = (7.71
\pm 0.07) + (-0.723 \pm 0.006) \times \log_{10}(\nu L_{\nu}(24
\micron) [L_{\odot}])$.  According to this fit, ULIRGs with rest-frame
24 $\micron$ luminosities above about $10^{12}$ L$_{\odot}$ have 6.2
$\micron$ PAH EQWs below 0.1--0.2 $\micron$.  However, very luminous
infrared sources with strong PAH emission (6.2 $\micron$ PAH EQW $>$
0.3-0.5 $\micron$) have recently been found in high-redshift samples
that were selected in the submillimeter, which is dominated by cool
dust emission \citep{Lutz05,MenendezDelmestre07,Valiante07}.  For
example, of the five IRS spectra of submillimeter sources presented by
\citet{MenendezDelmestre07}, two have the wavelength coverage to
measure the 6.2 $\micron$ PAH feature.  These two
submillimeter-selected ULIRGs have rest-frame EQWs of approximately
0.6 and 0.8 $\micron$, comparable to the strongest PAH emission
measured for local starburst galaxies.  Their positions are marked in
Figure \ref{fig:pahew_v_lum24_medians}, and they are clearly well
above the fit to the local ULIRGs.  Additional examples have been
found in samples selected at 160 $\micron$
\citep[MIPSJ142824;][]{Desai06} or a combination of 24 $\micron$ flux
and mid-infrared color \citep{Yan05,Sajina07}.  These are also plotted
in Figure \ref{fig:pahew_v_lum24_medians}.  However, sources with high
rest-frame 24 $\micron$ luminosities and large 6.2 $\micron$ PAH EQW
are rare in bright (f$_{24} > 0.7$ mJy) flux-limited samples selected
at 24 $\micron$ \citep[e.g.][]{Houck05,Weedman06}; these samples
appear dominated by AGN.  ULIRGs with high rest-frame 24 $\micron$
luminosities and strong PAH emission are absent from local samples,
yet they obviously exist at high redshift, and they can be found by
selecting on color, cold dust emission, or by probing to fainter flux
levels in the mid-infrared (observed $f_{24} < 0.3$ mJy).  At $z < 1$,
the generation of large amounts of warm dust seems to require the
presence of a dominant AGN.  The absence of extremely luminous
starburst-dominated ULIRGs at low redshift represents a real effect,
and not simply an observational bias.

\section{Summary and Conclusions}

Using a large sample of 107 ULIRGs observed with the IRS on board the
\textit{Spitzer Space Telescope}, we investigate the relationships
between the 6.2 and 11.3 $\micron$ PAH EQWs, continuum luminosity,
infrared spectral slope, optical spectroscopic classification, and
silicate optical depth.  Our results can be summarized as follows:

\begin{enumerate}

\item There is an extremely large range in detected 6.2 $\micron$ PAH
EQWs among ULIRGs, spanning more than two orders of magnitude from
$\sim$0.006--0.9 $\micron$.  ULIRGs classified as starburst-dominated
based on their far-infrared colors or optical spectra have median 6.2
$\micron$ PAH EQWs which are 50\% of that measured in low-luminosity
starburst galaxies.  An excess of hot dust, which is most prominent in
the far-infrared warm or Seyfert-like ULIRGs, appears to be the cause
of the decreased PAH EQW, as opposed to extinction or grain
destruction.  Approximately 55\% of LINER-like ULIRGs, 40\% of
HII-like ULIRGs, and 35\% of cold ULIRGs have very low 6.2 $\micron$
PAH EQWs, indicative of buried AGN or starbursts with supressed PAH
emission.  Most of these ULIRGs also have high apparent silicate
optical depths at 9.7 $\micron$.



\item The apparent silicate optical depth is not correlated with
far-infrared spectral slope.  The ULIRGs with the strongest silicate
absorption features have small 6.2 $\micron$ PAH EQWs and flat
spectral slopes in the far-infrared.  Their strong silicate absorption
implies large columns of cold dust, but their small PAH EQWs and
far-infrared colors imply substantial amounts of hot dust emission.
For these sources, a model in which a warm source is viewed through
either a smooth optically thick or clumpy dust shell seems
inconsistent with the data.



\item ULIRGs with the largest rest-frame 24 $\micron$ luminosities
have the smallest 6.2 $\micron$ PAH EQWs.  A fit to our sample
suggests $\log_{10}(6.2 \ \micron \ {\rm PAH \ EQW} [\micron]) = (7.71
\pm 0.07) + (-0.723 \pm 0.006) \times \log_{10}(\nu L_{\nu}(24
\micron) [L_{\odot}])$.  The range in PAH EQWs among luminous,
high-redshift sources observed with Spitzer is much larger than
implied by this correlation due to the presence of luminous sources
with very large 6.2 $\micron$ EQWs.  Extremely luminous sources of all
types are rare at low-redshift, but those that do exist seem to be AGN
dominated.

\end{enumerate}

\acknowledgements

We would like to thank R. Chary, M. Lacy, J. Surace, and A. Sajina for
insightful discussions.  Support for this work was provided by NASA
through an award issued by JPL/Caltech.

\bibliographystyle{apj}
\bibliography{references}

\begin{thebibliography}{66}
\expandafter\ifx\csname natexlab\endcsname\relax\def\natexlab#1{#1}\fi

\bibitem[{{Aitken} \& {Roche}(1985)}]{Aitken85}
{Aitken}, D.~K., \& {Roche}, P.~F. 1985, \mnras, 213, 777

\bibitem[{{Allen} {et~al.}(1991){Allen}, {Norris}, {Meadows}, \&
  {Roche}}]{Allen91}
{Allen}, D.~A., {Norris}, R.~P., {Meadows}, V.~S., \& {Roche}, P.~F. 1991,
  \mnras, 248, 528

\bibitem[{{Armus} {et~al.}(2006){Armus}, {Bernard-Salas}, {Spoon}, {Marshall},
  {Charmandaris}, {Higdon}, {Desai}, {Hao}, {Teplitz}, {Devost}, {Brandl},
  {Soifer}, \& {Houck}}]{Armus06}
{Armus}, L. {et~al.} 2006, \apj, 640, 204

\bibitem[{{Armus} {et~al.}(2007){Armus}, {Charmandaris}, {Bernard-Salas},
  {Spoon}, {Marshall}, {Higdon}, {Desai}, {Teplitz}, {Hao}, {Devost}, {Brandl},
  {Wu}, {Sloan}, {Soifer}, {Houck}, \& {Herter}}]{Armus07}
---. 2007, \apj, 656, 148

\bibitem[{{Armus} {et~al.}(2004){Armus}, {Charmandaris}, {Spoon}, {Houck},
  {Soifer}, {Brandl}, {Appleton}, {Teplitz}, {Higdon}, {Weedman}, {Devost},
  {Morris}, {Uchida}, {van Cleve}, {Barry}, {Sloan}, {Grillmair}, {Burgdorf},
  {Fajardo-Acosta}, {Ingalls}, {Higdon}, {Hao}, {Bernard-Salas}, {Herter},
  {Troeltzsch}, {Unruh}, \& {Winghart}}]{Armus04}
---. 2004, \apjs, 154, 178

\bibitem[{{Armus} {et~al.}(1987){Armus}, {Heckman}, \& {Miley}}]{Armus87}
{Armus}, L., {Heckman}, T., \& {Miley}, G. 1987, \aj, 94, 831

\bibitem[{{Armus} {et~al.}(1989){Armus}, {Heckman}, \& {Miley}}]{Armus89}
{Armus}, L., {Heckman}, T.~M., \& {Miley}, G.~K. 1989, \apj, 347, 727

\bibitem[{{Borys} {et~al.}(2006){Borys}, {Blain}, {Dey}, {Le Floc'h},
  {Jannuzi}, {Barnard}, {Bian}, {Brodwin}, {Men{\'e}ndez-Delmestre},
  {Thompson}, {Brand}, {Brown}, {Dowell}, {Eisenhardt}, {Farrah}, {Frayer},
  {Higdon}, {Higdon}, {Phillips}, {Soifer}, {Stern}, \& {Weedman}}]{Borys06}
{Borys}, C. {et~al.} 2006, \apj, 636, 134

\bibitem[{{Brandl} {et~al.}(2006){Brandl}, {Bernard-Salas}, {Spoon}, {Devost},
  {Sloan}, {Guilles}, {Wu}, {Houck}, {Weedman}, {Armus}, {Appleton}, {Soifer},
  {Charmandaris}, {Hao}, {Higdon}, \& {Herter}}]{Brandl06}
{Brandl}, B.~R. {et~al.} 2006, \apj, 653, 1129

\bibitem[{{Burston} {et~al.}(2001){Burston}, {Ward}, \& {Davies}}]{Burston01}
{Burston}, A.~J., {Ward}, M.~J., \& {Davies}, R.~I. 2001, \mnras, 326, 403

\bibitem[{{Cutri} {et~al.}(1994){Cutri}, {Huchra}, {Low}, {Brown}, \& {Vanden
  Bout}}]{Cutri94}
{Cutri}, R.~M., {Huchra}, J.~P., {Low}, F.~J., {Brown}, R.~L., \& {Vanden
  Bout}, P.~A. 1994, \apjl, 424, L65

\bibitem[{{Dannerbauer} {et~al.}(2005){Dannerbauer}, {Rigopoulou}, {Lutz},
  {Genzel}, {Sturm}, \& {Moorwood}}]{Dannerbauer05}
{Dannerbauer}, H., {Rigopoulou}, D., {Lutz}, D., {Genzel}, R., {Sturm}, E., \&
  {Moorwood}, A.~F.~M. 2005, \aap, 441, 999

\bibitem[{{Davies} {et~al.}(2003){Davies}, {Sternberg}, {Lehnert}, \&
  {Tacconi-Garman}}]{Davies03}
{Davies}, R.~I., {Sternberg}, A., {Lehnert}, M., \& {Tacconi-Garman}, L.~E.
  2003, \apj, 597, 907

\bibitem[{{de Grijp} {et~al.}(1985){de Grijp}, {Miley}, {Lub}, \& {de
  Jong}}]{deGrijp85}
{de Grijp}, M.~H.~K., {Miley}, G.~K., {Lub}, J., \& {de Jong}, T. 1985, \nat,
  314, 240

\bibitem[{{Desai} {et~al.}(2006){Desai}, {Armus}, {Soifer}, {Weedman},
  {Higdon}, {Bian}, {Borys}, {Spoon}, {Charmandaris}, {Brand}, {Brown}, {Dey},
  {Higdon}, {Houck}, {Jannuzi}, {Le Floc'h}, {Ashby}, \& {Smith}}]{Desai06}
{Desai}, V. {et~al.} 2006, \apj, 641, 133

\bibitem[{{Duc} {et~al.}(1997){Duc}, {Mirabel}, \& {Maza}}]{Duc97}
{Duc}, P.-A., {Mirabel}, I.~F., \& {Maza}, J. 1997, \aaps, 124, 533

\bibitem[{{Farrah} {et~al.}(2007){Farrah}, {Bernard-Salas}, {Spoon}, {Soifer},
  {Armus}, {Brandl}, {Charmandaris}, {Desai}, {Higdon}, {Devost}, \&
  {Houck}}]{Farrah07}
{Farrah}, D. {et~al.} 2007, ArXiv e-prints, 706

\bibitem[{{Franceschini} {et~al.}(2001){Franceschini}, {Aussel}, {Cesarsky},
  {Elbaz}, \& {Fadda}}]{Franceschini01}
{Franceschini}, A., {Aussel}, H., {Cesarsky}, C.~J., {Elbaz}, D., \& {Fadda},
  D. 2001, \aap, 378, 1

\bibitem[{{Genzel} {et~al.}(1998){Genzel}, {Lutz}, {Sturm}, {Egami}, {Kunze},
  {Moorwood}, {Rigopoulou}, {Spoon}, {Sternberg}, {Tacconi-Garman}, {Tacconi},
  \& {Thatte}}]{Genzel98}
{Genzel}, R. {et~al.} 1998, \apj, 498, 579

\bibitem[{{Goldader} {et~al.}(1995){Goldader}, {Joseph}, {Doyon}, \&
  {Sanders}}]{Goldader95}
{Goldader}, J.~D., {Joseph}, R.~D., {Doyon}, R., \& {Sanders}, D.~B. 1995,
  \apj, 444, 97

\bibitem[{{Hao} {et~al.}(2007){Hao}, {Weedman}, {Spoon}, {Marshall},
  {Levenson}, {Elitzur}, \& {Houck}}]{Hao07}
{Hao}, L., {Weedman}, D.~W., {Spoon}, H.~W.~W., {Marshall}, J.~A., {Levenson},
  N.~A., {Elitzur}, M., \& {Houck}, J.~R. 2007, \apjl, 655, L77

\bibitem[{{Higdon} {et~al.}(2004){Higdon}, {Devost}, {Higdon}, {Brandl},
  {Houck}, {Hall}, {Barry}, {Charmandaris}, {Smith}, {Sloan}, \&
  {Green}}]{Higdon04}
{Higdon}, S.~J.~U. {et~al.} 2004, \pasp, 116, 975

\bibitem[{{Houck} {et~al.}(2004){Houck}, {Roellig}, {van Cleve}, {Forrest},
  {Herter}, {Lawrence}, {Matthews}, {Reitsema}, {Soifer}, {Watson}, {Weedman},
  {Huisjen}, {Troeltzsch}, {Barry}, {Bernard-Salas}, {Blacken}, {Brandl},
  {Charmandaris}, {Devost}, {Gull}, {Hall}, {Henderson}, {Higdon}, {Pirger},
  {Schoenwald}, {Sloan}, {Uchida}, {Appleton}, {Armus}, {Burgdorf},
  {Fajardo-Acosta}, {Grillmair}, {Ingalls}, {Morris}, \& {Teplitz}}]{Houck04}
{Houck}, J.~R. {et~al.} 2004, \apjs, 154, 18

\bibitem[{{Houck} {et~al.}(2005){Houck}, {Soifer}, {Weedman}, {Higdon},
  {Higdon}, {Herter}, {Brown}, {Dey}, {Jannuzi}, {Le Floc'h}, {Rieke}, {Armus},
  {Charmandaris}, {Brandl}, \& {Teplitz}}]{Houck05}
---. 2005, \apjl, 622, L105

\bibitem[{{Imanishi} {et~al.}(2007){Imanishi}, {Dudley}, {Maiolino}, {Maloney},
  {Nakagawa}, \& {Risaliti}}]{Imanishi07}
{Imanishi}, M., {Dudley}, C.~C., {Maiolino}, R., {Maloney}, P.~R., {Nakagawa},
  T., \& {Risaliti}, G. 2007, \apjs, 171, 72

\bibitem[{{Kim} {et~al.}(1998){Kim}, {Veilleux}, \& {Sanders}}]{Kim98}
{Kim}, D.-C., {Veilleux}, S., \& {Sanders}, D.~B. 1998, \apj, 508, 627

\bibitem[{{Kleinmann} {et~al.}(1988){Kleinmann}, {Hamilton}, {Keel},
  {Wynn-Williams}, {Eales}, {Becklin}, \& {Kuntz}}]{Kleinmann88}
{Kleinmann}, S.~G., {Hamilton}, D., {Keel}, W.~C., {Wynn-Williams}, C.~G.,
  {Eales}, S.~A., {Becklin}, E.~E., \& {Kuntz}, K.~D. 1988, \apj, 328, 161

\bibitem[{{Kormendy} \& {Sanders}(1992)}]{Kormendy92}
{Kormendy}, J., \& {Sanders}, D.~B. 1992, \apjl, 390, L53

\bibitem[{{Laurent} {et~al.}(2000){Laurent}, {Mirabel}, {Charmandaris},
  {Gallais}, {Madden}, {Sauvage}, {Vigroux}, \& {Cesarsky}}]{Laurent00}
{Laurent}, O., {Mirabel}, I.~F., {Charmandaris}, V., {Gallais}, P., {Madden},
  S.~C., {Sauvage}, M., {Vigroux}, L., \& {Cesarsky}, C. 2000, \aap, 359, 887

\bibitem[{{Le Floc'h} {et~al.}(2005){Le Floc'h}, {Papovich}, {Dole}, {Bell},
  {Lagache}, {Rieke}, {Egami}, {P{\'e}rez-Gonz{\'a}lez}, {Alonso-Herrero},
  {Rieke}, {Blaylock}, {Engelbracht}, {Gordon}, {Hines}, {Misselt}, {Morrison},
  \& {Mould}}]{LeFloch05}
{Le Floc'h}, E. {et~al.} 2005, \apj, 632, 169

\bibitem[{{Levenson} {et~al.}(2007){Levenson}, {Sirocky}, {Hao}, {Spoon},
  {Marshall}, {Elitzur}, \& {Houck}}]{Levenson07}
{Levenson}, N.~A., {Sirocky}, M.~M., {Hao}, L., {Spoon}, H.~W.~W., {Marshall},
  J.~A., {Elitzur}, M., \& {Houck}, J.~R. 2007, \apjl, 654, L45

\bibitem[{{Lutz} {et~al.}(1998){Lutz}, {Spoon}, {Rigopoulou}, {Moorwood}, \&
  {Genzel}}]{Lutz98}
{Lutz}, D., {Spoon}, H.~W.~W., {Rigopoulou}, D., {Moorwood}, A.~F.~M., \&
  {Genzel}, R. 1998, \apjl, 505, L103

\bibitem[{{Lutz} {et~al.}(2005){Lutz}, {Valiante}, {Sturm}, {Genzel},
  {Tacconi}, {Lehnert}, {Sternberg}, \& {Baker}}]{Lutz05}
{Lutz}, D., {Valiante}, E., {Sturm}, E., {Genzel}, R., {Tacconi}, L.~J.,
  {Lehnert}, M.~D., {Sternberg}, A., \& {Baker}, A.~J. 2005, \apjl, 625, L83

\bibitem[{{Lutz} {et~al.}(1999){Lutz}, {Veilleux}, \& {Genzel}}]{Lutz99}
{Lutz}, D., {Veilleux}, S., \& {Genzel}, R. 1999, \apjl, 517, L13

\bibitem[{{Men{\'e}ndez-Delmestre} {et~al.}(2007){Men{\'e}ndez-Delmestre},
  {Blain}, {Alexander}, {Smail}, {Armus}, {Chapman}, {Frayer}, {Ivison}, \&
  {Teplitz}}]{MenendezDelmestre07}
{Men{\'e}ndez-Delmestre}, K. {et~al.} 2007, \apjl, 655, L65

\bibitem[{{Moshir}(1990)}]{Moshir90}
{Moshir}, M. 1990, in IRAS Faint Source Catalogue, version 2.0 (1990), 0--+

\bibitem[{{Murphy} {et~al.}(2001){Murphy}, {Soifer}, {Matthews}, {Armus}, \&
  {Kiger}}]{Murphy01}
{Murphy}, Jr., T.~W., {Soifer}, B.~T., {Matthews}, K., {Armus}, L., \& {Kiger},
  J.~R. 2001, \aj, 121, 97

\bibitem[{{Murphy} {et~al.}(1999){Murphy}, {Soifer}, {Matthews}, {Kiger}, \&
  {Armus}}]{Murphy99}
{Murphy}, Jr., T.~W., {Soifer}, B.~T., {Matthews}, K., {Kiger}, J.~R., \&
  {Armus}, L. 1999, \apjl, 525, L85

\bibitem[{{Osterbrock} \& {De Robertis}(1985)}]{Osterbrock85}
{Osterbrock}, D.~E., \& {De Robertis}, M.~M. 1985, \pasp, 97, 1129

\bibitem[{{P{\'e}rez-Gonz{\'a}lez} {et~al.}(2005){P{\'e}rez-Gonz{\'a}lez},
  {Rieke}, {Egami}, {Alonso-Herrero}, {Dole}, {Papovich}, {Blaylock}, {Jones},
  {Rieke}, {Rigby}, {Barmby}, {Fazio}, {Huang}, \& {Martin}}]{PerezGonzalez05}
{P{\'e}rez-Gonz{\'a}lez}, P.~G. {et~al.} 2005, \apj, 630, 82

\bibitem[{{Rigopoulou} {et~al.}(1999){Rigopoulou}, {Spoon}, {Genzel}, {Lutz},
  {Moorwood}, \& {Tran}}]{Rigopoulou99}
{Rigopoulou}, D., {Spoon}, H.~W.~W., {Genzel}, R., {Lutz}, D., {Moorwood},
  A.~F.~M., \& {Tran}, Q.~D. 1999, \aj, 118, 2625

\bibitem[{{Roussel} {et~al.}(2006){Roussel}, {Helou}, {Smith}, {Draine},
  {Hollenbach}, {Moustakas}, {Spoon}, {Kennicutt}, {Rieke}, {Walter}, {Armus},
  {Dale}, {Sheth}, {Bendo}, {Engelbracht}, {Gordon}, {Meyer}, {Regan}, \&
  {Murphy}}]{Roussel06}
{Roussel}, H. {et~al.} 2006, \apj, 646, 841

\bibitem[{{Sajina} {et~al.}(2007){Sajina}, {Yan}, {Armus}, {Choi}, {Fadda},
  {Helou}, \& {Spoon}}]{Sajina07}
{Sajina}, A., {Yan}, L., {Armus}, L., {Choi}, P., {Fadda}, D., {Helou}, G., \&
  {Spoon}, H. 2007, ArXiv e-prints, 704

\bibitem[{{Sanders} {et~al.}(2003){Sanders}, {Mazzarella}, {Kim}, {Surace}, \&
  {Soifer}}]{Sanders03}
{Sanders}, D.~B., {Mazzarella}, J.~M., {Kim}, D.-C., {Surace}, J.~A., \&
  {Soifer}, B.~T. 2003, \aj, 126, 1607

\bibitem[{{Sanders} \& {Mirabel}(1996)}]{SandersMirabel96}
{Sanders}, D.~B., \& {Mirabel}, I.~F. 1996, \araa, 34, 749

\bibitem[{{Sanders} {et~al.}(1988{\natexlab{a}}){Sanders}, {Soifer}, {Elias},
  {Madore}, {Matthews}, {Neugebauer}, \& {Scoville}}]{Sanders88}
{Sanders}, D.~B., {Soifer}, B.~T., {Elias}, J.~H., {Madore}, B.~F., {Matthews},
  K., {Neugebauer}, G., \& {Scoville}, N.~Z. 1988{\natexlab{a}}, \apj, 325, 74

\bibitem[{{Sanders} {et~al.}(1988{\natexlab{b}}){Sanders}, {Soifer}, {Elias},
  {Neugebauer}, \& {Matthews}}]{Sanders88b}
{Sanders}, D.~B., {Soifer}, B.~T., {Elias}, J.~H., {Neugebauer}, G., \&
  {Matthews}, K. 1988{\natexlab{b}}, \apjl, 328, L35

\bibitem[{{Schweitzer} {et~al.}(2006){Schweitzer}, {Lutz}, {Sturm}, {Contursi},
  {Tacconi}, {Lehnert}, {Dasyra}, {Genzel}, {Veilleux}, {Rupke}, {Kim},
  {Baker}, {Netzer}, {Sternberg}, {Mazzarella}, \& {Lord}}]{Schweitzer06}
{Schweitzer}, M. {et~al.} 2006, \apj, 649, 79

\bibitem[{{Scott} {et~al.}(2002){Scott}, {Fox}, {Dunlop}, {Serjeant},
  {Peacock}, {Ivison}, {Oliver}, {Mann}, {Lawrence}, {Efstathiou},
  {Rowan-Robinson}, {Hughes}, {Archibald}, {Blain}, \& {Longair}}]{Scott02}
{Scott}, S.~E. {et~al.} 2002, \mnras, 331, 817

\bibitem[{{Soifer} {et~al.}(2000){Soifer}, {Neugebauer}, {Matthews}, {Egami},
  {Becklin}, {Weinberger}, {Ressler}, {Werner}, {Evans}, {Scoville}, {Surace},
  \& {Condon}}]{Soifer00}
{Soifer}, B.~T. {et~al.} 2000, \aj, 119, 509

\bibitem[{{Soifer} {et~al.}(1987){Soifer}, {Sanders}, {Madore}, {Neugebauer},
  {Danielson}, {Elias}, {Lonsdale}, \& {Rice}}]{Soifer87}
{Soifer}, B.~T., {Sanders}, D.~B., {Madore}, B.~F., {Neugebauer}, G.,
  {Danielson}, G.~E., {Elias}, J.~H., {Lonsdale}, C.~J., \& {Rice}, W.~L. 1987,
  \apj, 320, 238

\bibitem[{{Spoon} {et~al.}(2002){Spoon}, {Keane}, {Tielens}, {Lutz},
  {Moorwood}, \& {Laurent}}]{Spoon02}
{Spoon}, H.~W.~W., {Keane}, J.~V., {Tielens}, A.~G.~G.~M., {Lutz}, D.,
  {Moorwood}, A.~F.~M., \& {Laurent}, O. 2002, \aap, 385, 1022

\bibitem[{{Spoon} {et~al.}(2007){Spoon}, {Marshall}, {Houck}, {Elitzur}, {Hao},
  {Armus}, {Brandl}, \& {Charmandaris}}]{Spoon07}
{Spoon}, H.~W.~W., {Marshall}, J.~A., {Houck}, J.~R., {Elitzur}, M., {Hao}, L.,
  {Armus}, L., {Brandl}, B.~R., \& {Charmandaris}, V. 2007, \apjl, 654, L49

\bibitem[{{Stanford} {et~al.}(2000){Stanford}, {Stern}, {van Breugel}, \& {De
  Breuck}}]{Stanford00}
{Stanford}, S.~A., {Stern}, D., {van Breugel}, W., \& {De Breuck}, C. 2000,
  \apjs, 131, 185

\bibitem[{{Strauss} {et~al.}(1992){Strauss}, {Huchra}, {Davis}, {Yahil},
  {Fisher}, \& {Tonry}}]{Strauss92}
{Strauss}, M.~A., {Huchra}, J.~P., {Davis}, M., {Yahil}, A., {Fisher}, K.~B.,
  \& {Tonry}, J. 1992, \apjs, 83, 29

\bibitem[{{Sturm} {et~al.}(2002){Sturm}, {Lutz}, {Verma}, {Netzer},
  {Sternberg}, {Moorwood}, {Oliva}, \& {Genzel}}]{Sturm02}
{Sturm}, E., {Lutz}, D., {Verma}, A., {Netzer}, H., {Sternberg}, A.,
  {Moorwood}, A.~F.~M., {Oliva}, E., \& {Genzel}, R. 2002, \aap, 393, 821

\bibitem[{{Sturm} {et~al.}(2006){Sturm}, {Rupke}, {Contursi}, {Kim}, {Lutz},
  {Netzer}, {Veilleux}, {Genzel}, {Lehnert}, {Tacconi}, {Maoz}, {Mazzarella},
  {Lord}, {Sanders}, \& {Sternberg}}]{Sturm06}
{Sturm}, E. {et~al.} 2006, \apjl, 653, L13

\bibitem[{{Tran} {et~al.}(2001){Tran}, {Lutz}, {Genzel}, {Rigopoulou}, {Spoon},
  {Sturm}, {Gerin}, {Hines}, {Moorwood}, {Sanders}, {Scoville}, {Taniguchi}, \&
  {Ward}}]{Tran01}
{Tran}, Q.~D. {et~al.} 2001, \apj, 552, 527

\bibitem[{{Valiante} {et~al.}(2007){Valiante}, {Lutz}, {Sturm}, {Genzel},
  {Tacconi}, {Lehnert}, \& {Baker}}]{Valiante07}
{Valiante}, E., {Lutz}, D., {Sturm}, E., {Genzel}, R., {Tacconi}, L.~J.,
  {Lehnert}, M.~D., \& {Baker}, A.~J. 2007, \apj, 660, 1060

\bibitem[{{Veilleux} {et~al.}(1999{\natexlab{a}}){Veilleux}, {Kim}, \&
  {Sanders}}]{Veilleux99}
{Veilleux}, S., {Kim}, D.-C., \& {Sanders}, D.~B. 1999{\natexlab{a}}, \apj,
  522, 113

\bibitem[{{Veilleux} {et~al.}(1995){Veilleux}, {Kim}, {Sanders}, {Mazzarella},
  \& {Soifer}}]{Veilleux95}
{Veilleux}, S., {Kim}, D.-C., {Sanders}, D.~B., {Mazzarella}, J.~M., \&
  {Soifer}, B.~T. 1995, \apjs, 98, 171

\bibitem[{{Veilleux} {et~al.}(1997){Veilleux}, {Sanders}, \&
  {Kim}}]{Veilleux97}
{Veilleux}, S., {Sanders}, D.~B., \& {Kim}, D.-C. 1997, \apj, 484, 92

\bibitem[{{Veilleux} {et~al.}(1999{\natexlab{b}}){Veilleux}, {Sanders}, \&
  {Kim}}]{Veilleux99b}
---. 1999{\natexlab{b}}, \apj, 522, 139

\bibitem[{{Voit}(1992)}]{Voit92}
{Voit}, G.~M. 1992, \mnras, 258, 841

\bibitem[{{Weedman} {et~al.}(2006){Weedman}, {Soifer}, {Hao}, {Higdon},
  {Higdon}, {Houck}, {Le Floc'h}, {Brown}, {Dey}, {Jannuzi}, {Rieke}, {Desai},
  {Bian}, {Thompson}, {Armus}, {Teplitz}, {Eisenhardt}, \&
  {Willner}}]{Weedman06}
{Weedman}, D.~W. {et~al.} 2006, \apj, 651, 101

\bibitem[{{Yan} {et~al.}(2005){Yan}, {Chary}, {Armus}, {Teplitz}, {Helou},
  {Frayer}, {Fadda}, {Surace}, \& {Choi}}]{Yan05}
{Yan}, L. {et~al.} 2005, \apj, 628, 604

\end{thebibliography}

\clearpage

\begin{deluxetable}{cccccccccc}
  \tabletypesize{\scriptsize}
  \tablecaption{6.2 and 11.3 $\micron$ PAH EQWs versus spectral class}
  \tablecolumns{10}
  \tablewidth{0pc}
  \tablehead{
    \colhead{} & 
    \colhead{} & 
    \multicolumn{2}{c}{uncorrected $6.2\mu {\rm m}$ PAH EQW} &
    \colhead{} &
    \multicolumn{2}{c}{corrected $6.2\mu {\rm m}$ PAH EQW} &  
    \colhead{} & 
    \multicolumn{2}{c}{11.3 $\micron$ PAH EQW}\\
    \cline{3-4} \cline{6-7} \cline{9-10} \\
    \colhead{spectral class} &
    \colhead{number}   &
    \colhead{median} &
    \colhead{average} &
    \colhead{} &
    \colhead{median} &
    \colhead{average} & 
    \colhead{} &
    \colhead{median} &
    \colhead{average} \\
    \colhead{} &
    \colhead{} &
    \colhead{($\mu$m)} &
    \colhead{($\mu$m)} &
    \colhead{} &
    \colhead{($\micron$)} &
    \colhead{($\micron$)} &
    \colhead{} &
    \colhead{($\micron$)} &
    \colhead{($\micron$)}
}

\startdata
HII   & 15 & 0.32 & 0.30 $\pm$ 0.16 & & 0.28 & 0.27 $\pm$ 0.18 & & 0.39 & 0.41 $\pm$ 0.20\\
LINER & 22 & 0.19 & 0.27 $\pm$ 0.20 & & 0.11 & 0.17 $\pm$ 0.14 & & 0.50 & 0.46 $\pm$ 0.18\\
S1    &  6 & 0.02 & 0.04 $\pm$ 0.05 & & 0.02 & 0.04 $\pm$ 0.05 & & 0.03 & 0.03 $\pm$ 0.03\\
S2    & 21 & 0.05 & 0.10 $\pm$ 0.11 & & 0.05 & 0.08 $\pm$ 0.07 & & 0.09 & 0.16 $\pm$ 0.15\\
S1+S2 & 27 & 0.05 & 0.09 $\pm$ 0.10 & & 0.04 & 0.07 $\pm$ 0.07 & & 0.07 & 0.13 $\pm$ 0.14\\
      &    & & & &      &                 & &      &                \\
Cold  & 66 & 0.30 & 0.32 $\pm$ 0.18 & & 0.24 & 0.27 $\pm$ 0.18 & & 0.50 & 0.52 $\pm$ 0.24\\
Warm  & 41 & 0.04 & 0.08 $\pm$ 0.09 & & 0.04 & 0.06 $\pm$ 0.08 & & 0.07 & 0.13 $\pm$ 0.13\\
\enddata

\tablecomments{Statistics on the 6.2 and 11.3 $\micron$ PAH EQWs of
the ULIRGs in our sample, broken down by spectral class (Column 1).
Column 2 gives the number of ULIRGs in each class.  Columns 3 and 4
show the median and average 6.2 $\micron$ PAH EQWs, respectively,
uncorrected for ice absorption.  Columns 5 and 6 give the same
statistics for the ice-corrected 6.2 $\micron$ PAH EQW (see
\citet{Spoon07} for details on this correction).  Column 7 shows the
median and average 11.3 $\micron$ PAH EQWs, uncorrected for the
underlying silicate absorption.  All medians and averages were
computed by including sources with upper limits as though they were
detections.  Optical classifications are from
\citet{Duc97,Veilleux97,Veilleux95,Cutri94,Allen91,Armus89,Kleinmann88,Sanders88b}.
The far-infrared classifications of cold ($f_{\nu}(25
\micron)/f_{\nu}(60 \micron) < 0.2$) and warm ($f_{\nu}(25
\micron)/f_{\nu}(60 \micron) \ge 0.2$) are based on the rest-frame 25
and 60 $\micron$ values calculated as described in
\S{\ref{sec:Observations}}.}

\label{table:pahew_v_class}
\end{deluxetable}

\begin{deluxetable}{ccccccccccc}
  \tabletypesize{\scriptsize}
  \tablecaption{6.2 and 11.3 $\micron$ PAH EQW versus 24 $\micron$ luminosity}
  \tablecolumns{11}
  \tablewidth{0pc}
  \tablehead{
    \multicolumn{2}{c}{24 $\micron$}& \colhead{} & \multicolumn{2}{c}{uncorrected 6.2 $\micron$ PAH EQW} & \colhead{} & \multicolumn{2}{c}{corrected 6.2 $\micron$ PAH EQW} & \colhead{} & \multicolumn{2}{c}{11.3 $\micron$ PAH EQW}\\
    \cline{1-2} \cline{4-5} \cline{7-8} \cline{10-11}\\
    \colhead{Class} &
    \colhead{Luminosity}   &
    \colhead{} &
    \colhead{median} &
    \colhead{average} &
    \colhead{} &
    \colhead{median} &
    \colhead{average} & 
    \colhead{} & 
    \colhead{median} & 
    \colhead{average}\\
    \colhead{} &
    \colhead{($10^{12}$L$_\odot$)} &
    \colhead{} &
    \colhead{($\mu$m)} &
    \colhead{($\mu$m)} &
    \colhead{} &
    \colhead{($\mu$m)} &
    \colhead{($\mu$m)} & 
    \colhead{} &
    \colhead{($\mu$m)} &
    \colhead{($\mu$m)}
}
\startdata
1 & $0.03-0.20$ & & 0.43 & 0.47 $\pm$ 0.16 & & 0.41 & 0.41 $\pm$ 0.18 & & 0.58 & 0.61 $\pm$ 0.20\\
2 & $0.20-0.30$ & & 0.33 & 0.34 $\pm$ 0.18 & & 0.29 & 0.30 $\pm$ 0.17 & & 0.56 & 0.52 $\pm$ 0.24\\
3 & $0.30-0.44$ & & 0.16 & 0.23 $\pm$ 0.17 & & 0.10 & 0.17 $\pm$ 0.13 & & 0.44 & 0.39 $\pm$ 0.18\\
4 & $0.44-0.63$ & & 0.14 & 0.17 $\pm$ 0.11 & & 0.12 & 0.15 $\pm$ 0.11 & & 0.31 & 0.30 $\pm$ 0.15\\
5 & $0.63-1.15$ & & 0.10 & 0.12 $\pm$ 0.10 & & 0.06 & 0.07 $\pm$ 0.05 & & 0.19 & 0.31 $\pm$ 0.40\\
6 & $1.15-8.00$ & & 0.03 & 0.04 $\pm$ 0.05 & & 0.02 & 0.04 $\pm$ 0.05 & & 0.05 & 0.08 $\pm$ 0.08\\
\enddata

\tablecomments{Same as Table \ref{table:pahew_v_class}, except the
ULIRG sample is broken down into luminosity bins, rather than spectral
class.  Each bin contains 17-18 ULIRGs.}
\label{table:pahew_v_lum}
\end{deluxetable}


\begin{figure}
\centerline{
\plottwo{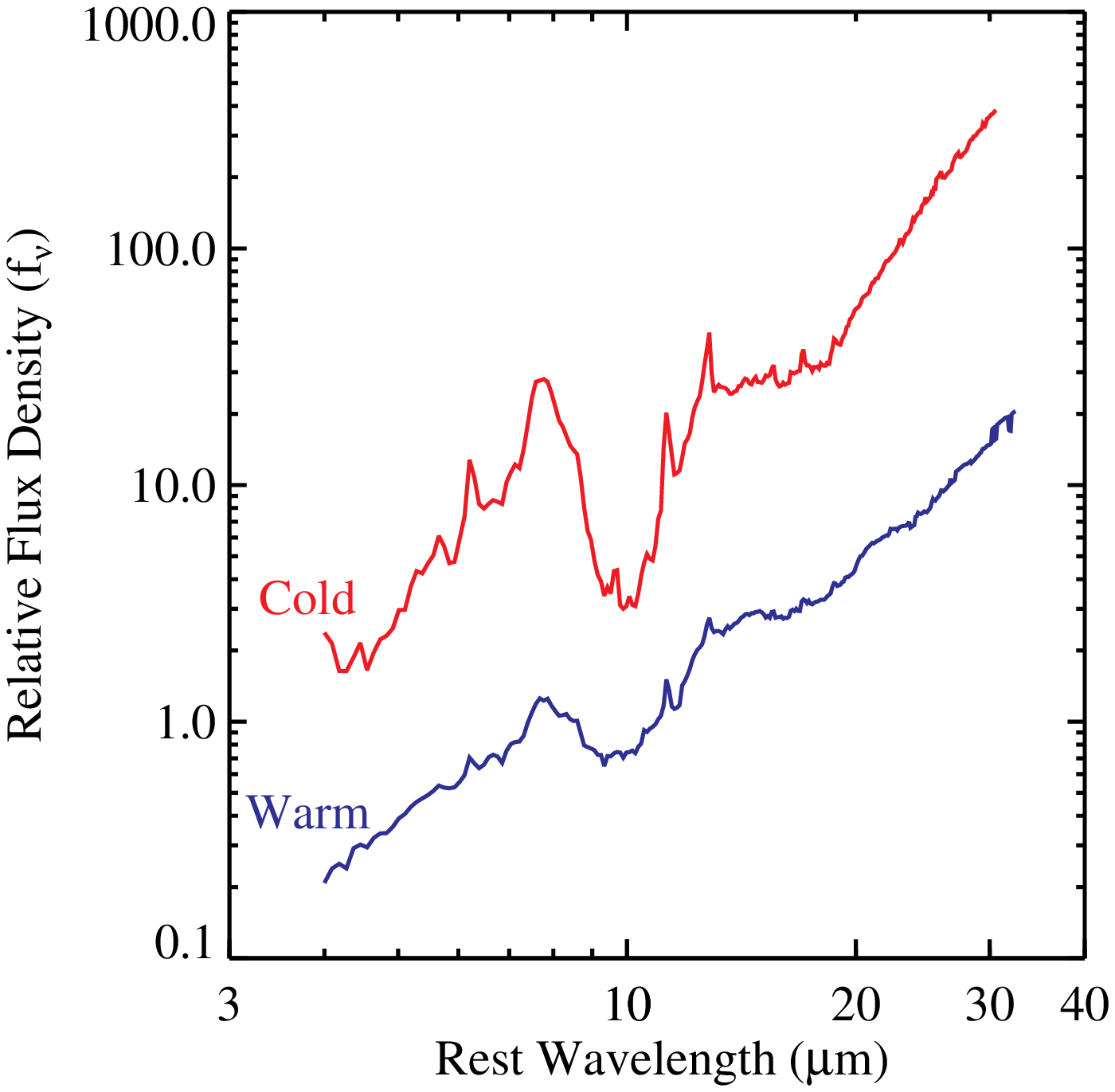}{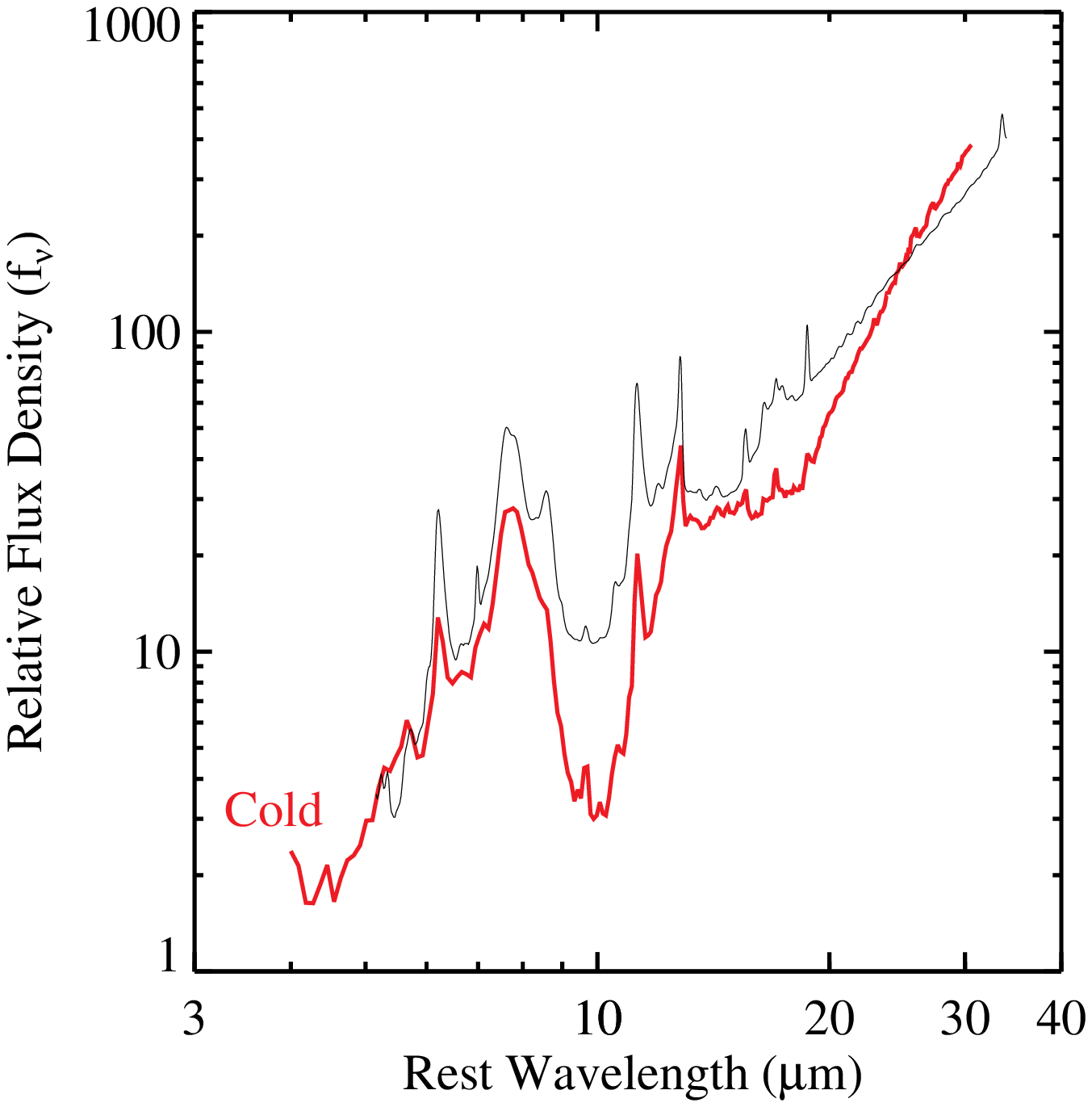}
}

\caption{\textit{Left:} The median IRS spectra of ULIRGs classified as
cold (red) and warm (blue), arbitrarily normalized for display
purposes. \textit{Right:} The median spectrum of cold ULIRGs (red)
compared to the average starburst spectrum from \citet{Brandl06}
(black), both normalized at 24 $\micron$.}

\label{fig:medianspectra_warmcold}
\end{figure}

\begin{figure}
\centerline{
\includegraphics[scale=1.0]{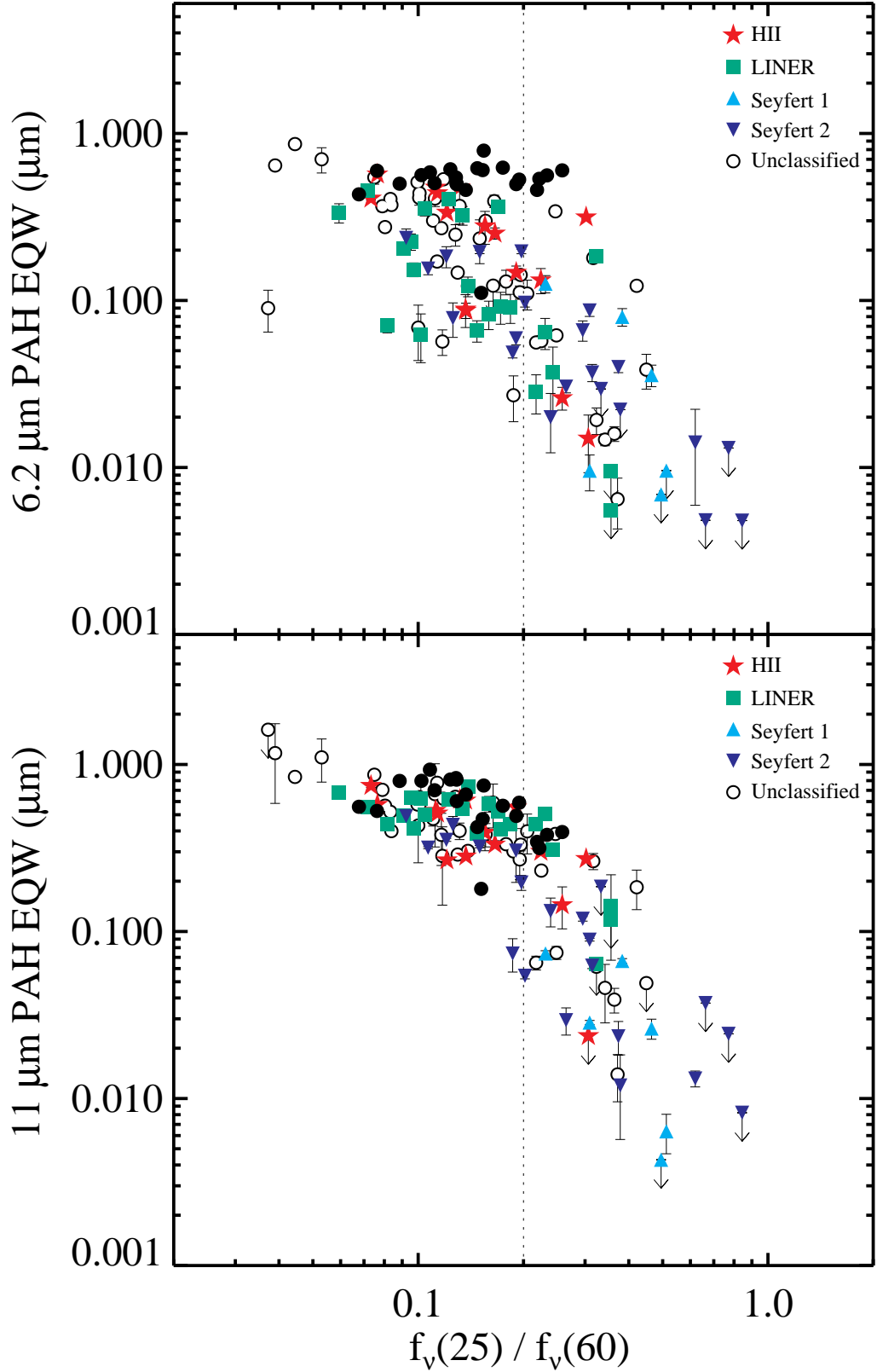}
}
\caption{The 6.2 $\micron$ (top) and 11.3 $\micron$ (bottom) PAH EQW
  versus spectral slope.  The ULIRGs are color-coded by optical
  spectroscopic classification: ULIRGs with HII-like optical spectra
  are plotted as red stars, LINER-like ULIRGs are shown as green
  squares, ULIRGs with Seyfert 1 optical spectra are shown as light
  blue triangles, ULIRGs with Seyfert 2 optical spectra are shown as
  dark blue upside-down triangles, and sources with unknown optical
  spectroscopic classifications are shown as white circles.  The black
  dots represent the starburst galaxies from \citet{Brandl06}.  The
  dotted vertical line represents the division between warm and cold
  sources adopted for this paper ($f_{\nu}(25) / f_{\nu}(60) = 0.2$),
  where cold sources lie to the left of the line and warm sources lie
  to the right.}

\label{fig:pah_v_slope}
\end{figure}

\begin{figure}
\centerline{
\includegraphics[scale=0.6]{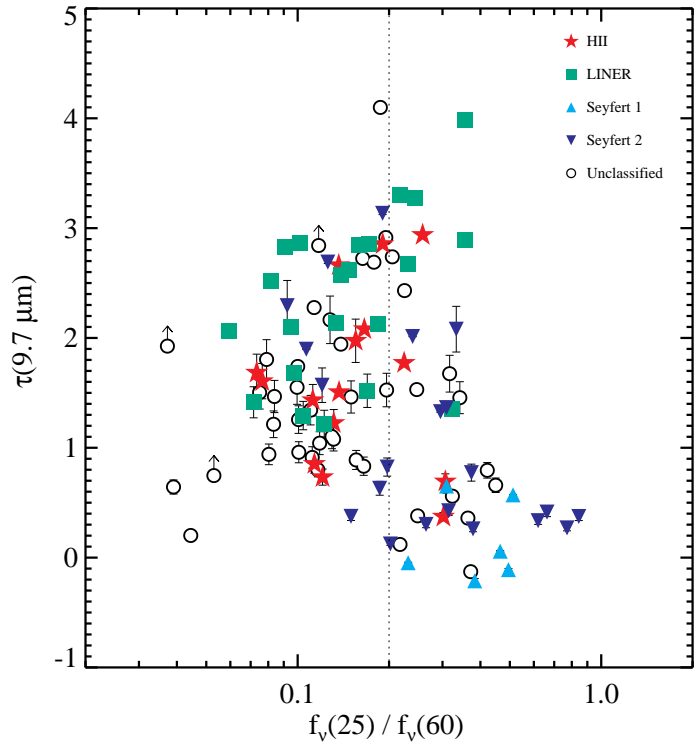}
}

\caption{The 9.7 $\micron$ silicate optical depth versus the
  mid-infrared spectral slope.  The ULIRGs are color-coded by optical
  spectroscopic classification, as in Figure \ref{fig:pah_v_slope}.
  The vertical dotted line is $f_{\nu}(25) / f_{\nu}(60) = 0.2$, the
  division between warm and cold sources adopted for this paper.  The
  spectral slope for the ULIRGs was computed from the IRS spectra as
  described in \S{\ref{sec:Observations}} and the spectral slope for
  the starbursts was computed from IRAS fluxes taken from
  \citet{Brandl06}.}

\label{fig:tau_v_slope}
\end{figure}


\begin{figure}
\centerline{
\plottwo{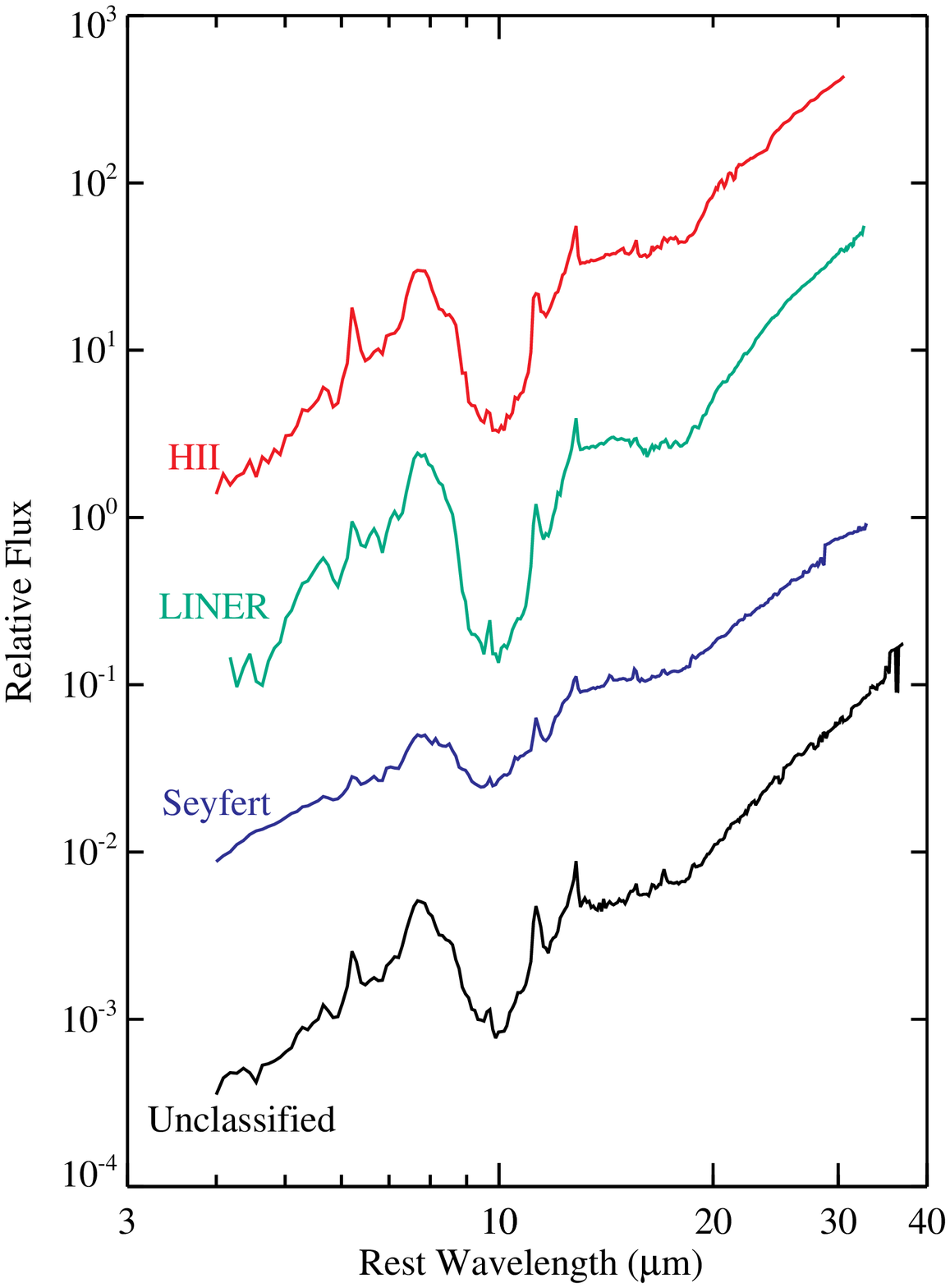}{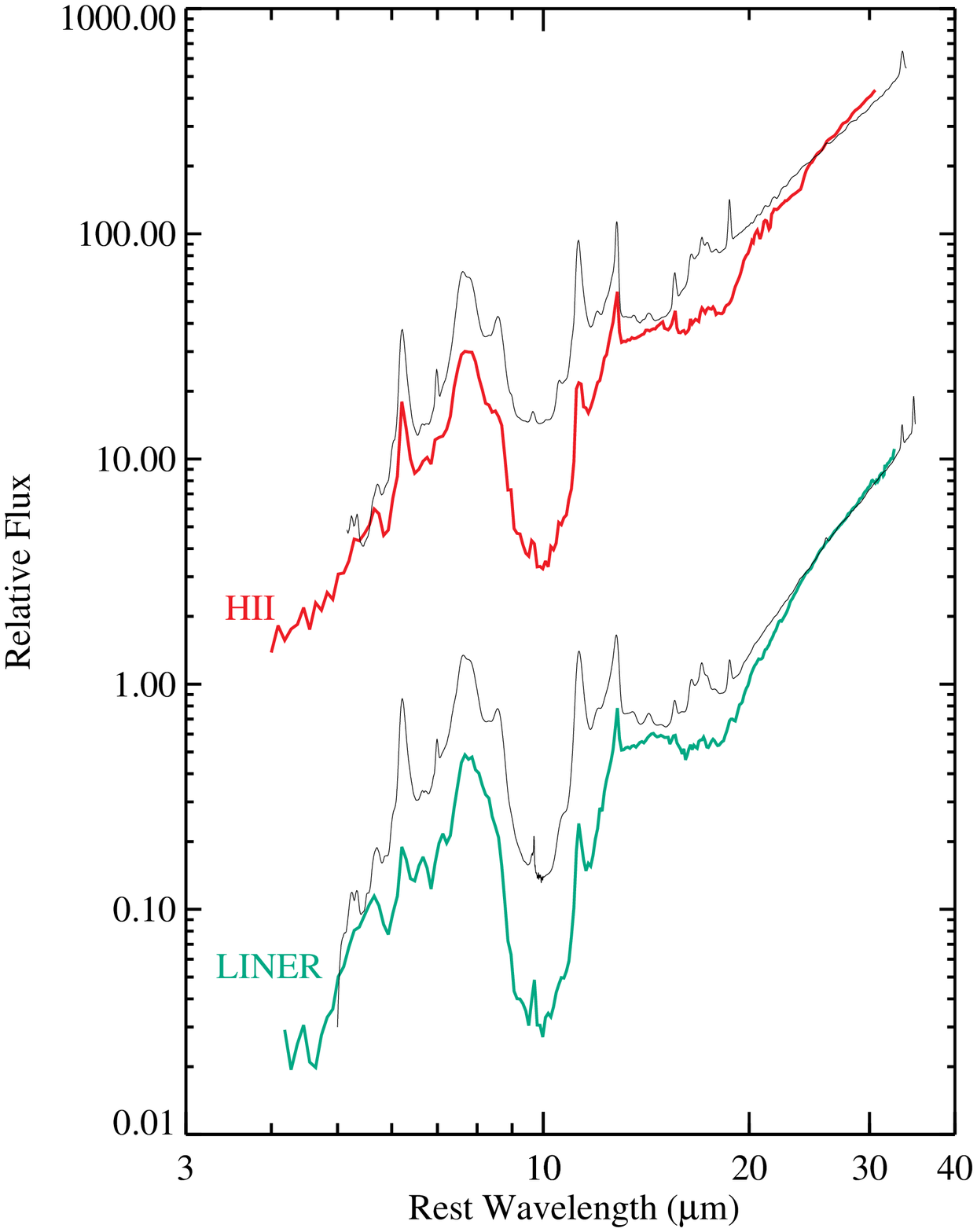}
}

\caption{\textit{Left:} The colored lines show the median IRS spectra
of ULIRGs optically classified as HII-like (red), LINER-like (green),
and Seyfert-like (blue).  The black line shows the median spectrum of
ULIRGs lacking optical classifications.  The normalizations are
arbitrary.  \textit{Right:} The median spectrum of HII-like ULIRGs
compared to the average starburst spectrum from \citet{Brandl06} (top
red and black lines respectively) and the median spectrum of
LINER-like ULIRGs compared to the average infrared-bright LINER
spectrum from \citet{Sturm06} (bottom green and black lines
respectively).  For wavelengths longer than 10 $\micron$, the
high-resolution \citet{Sturm06} spectrum was convolved with a Gaussian
profile with a FWHM of 0.2 $\micron$ to approximately match the
resolution of the low-resolution ULIRG spectrum.  The comparison
spectra were normalized to match the HII- and LINER-like ULIRGs at 24
$\micron$.}

\label{fig:medianspectra_optical}
\end{figure}


\begin{figure}
\includegraphics[scale=1.0]{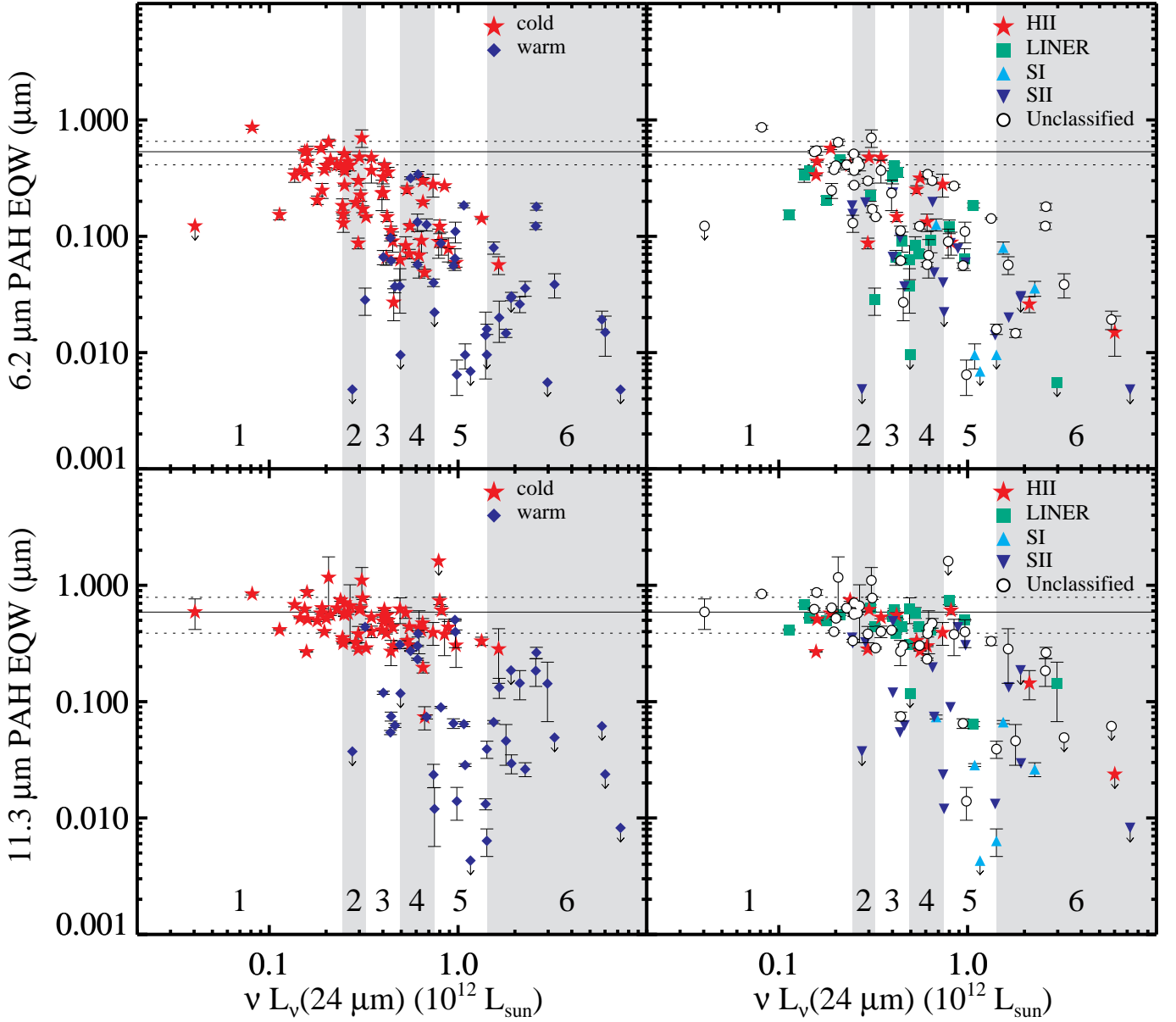}

\caption{The 6.2 $\micron$ (top row) and 11.3 $\micron$ (bottom row)
  PAH EQWs versus 24 $\micron$ rest-frame luminosity.  The ULIRG
  sample is color-coded by warm ($f_{\nu}(25 \micron)/f_{\nu}(60
  \micron) \ge 0.2$) and cold ($f_{\nu}(25 \micron)/f_{\nu}(60
  \micron) < 0.2$) mid-infrared spectral slope (left) and by optical
  spectroscopic classification as in Figure \ref{fig:tau_v_slope}
  (right).  The average PAH EQWs of starburst galaxies from
  \citet{Brandl06} are indicated by solid horizontal lines and the
  $\pm$1 $\sigma$ dispersions in these averages are shown by dotted
  horizontal lines.  Upper limits ($3\sigma$) to the PAH EQWs are
  indicated by downward pointing arrows. The six luminosity classes
  (1-6; see text and Table \ref{table:pahew_v_lum}) are labeled on
  the bottom of each panel and are alternately shaded for clarity.}

\label{fig:pahew_v_lum24}
\end{figure}


\begin{figure}
\centerline{ \includegraphics[scale=0.8]{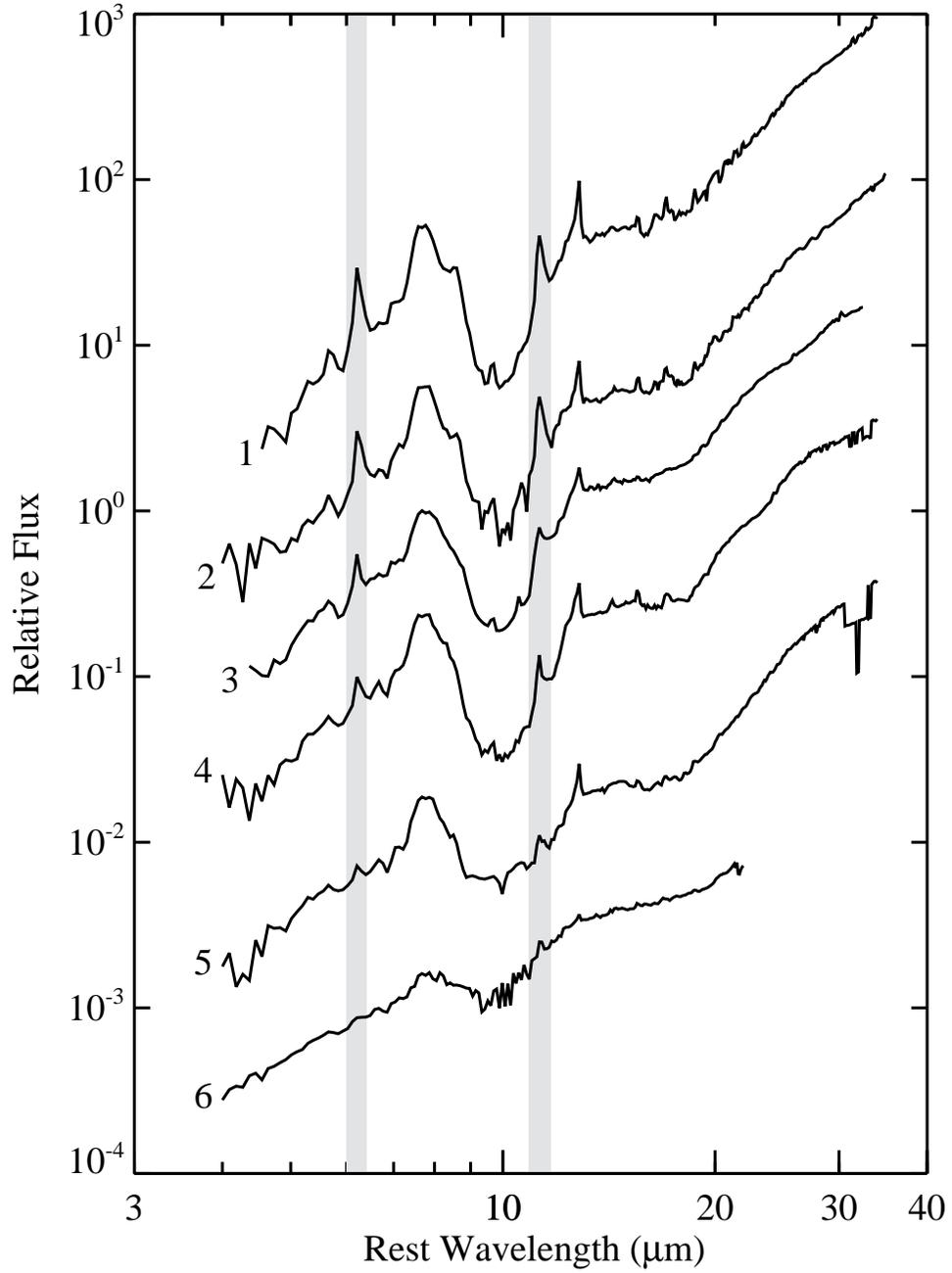}}

\caption{Median IRS ULIRG spectra for the six 24 $\micron$ rest-frame
luminosity bins shown in Figure \ref{fig:pahew_v_lum24}, arranged such
that the luminosity increases from top to bottom. The grey bands
highlight the positions of the 6.2 and 11.3 $\micron$ PAH features.}

\label{fig:medianspectra}
\end{figure}


\begin{figure}
\centerline{
\includegraphics[scale=1.0]{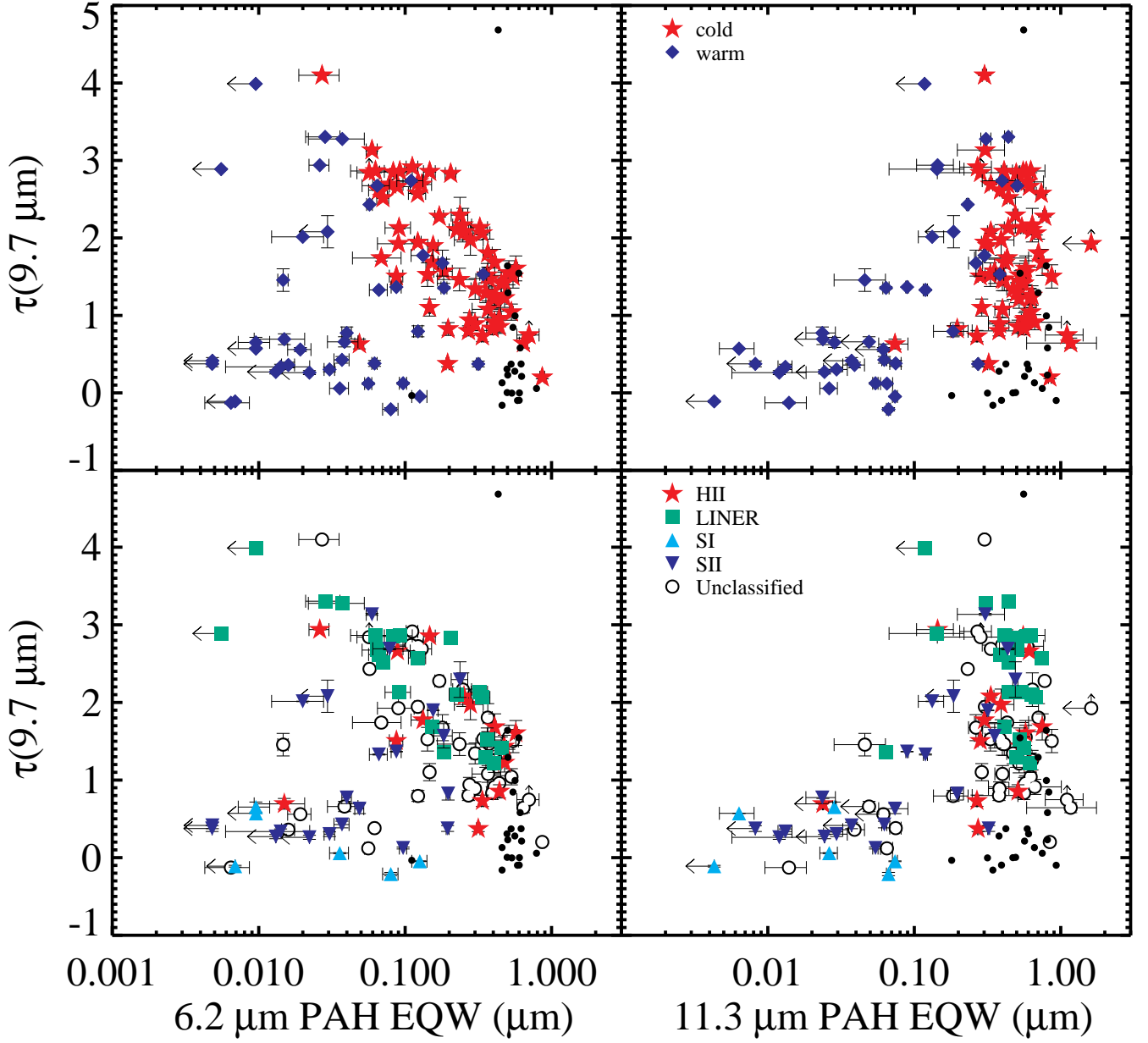}}

\caption{The strength of the 9.7 $\micron$ silicate absorption feature
versus the 6.2 (left) and 11.3 (right) $\micron$ PAH EQWs.  The ULIRGs
are color-coded according to spectral slope and optical classification
in the left and right panels, respectively.  The colored symbols are
identical to those in Figure \ref{fig:pahew_v_lum24}.  The small black
dots represent the starbursts from \citet{Brandl06}.}

\label{fig:fork}

\end{figure}


\begin{figure}
\centerline{
\includegraphics[scale=1.0]{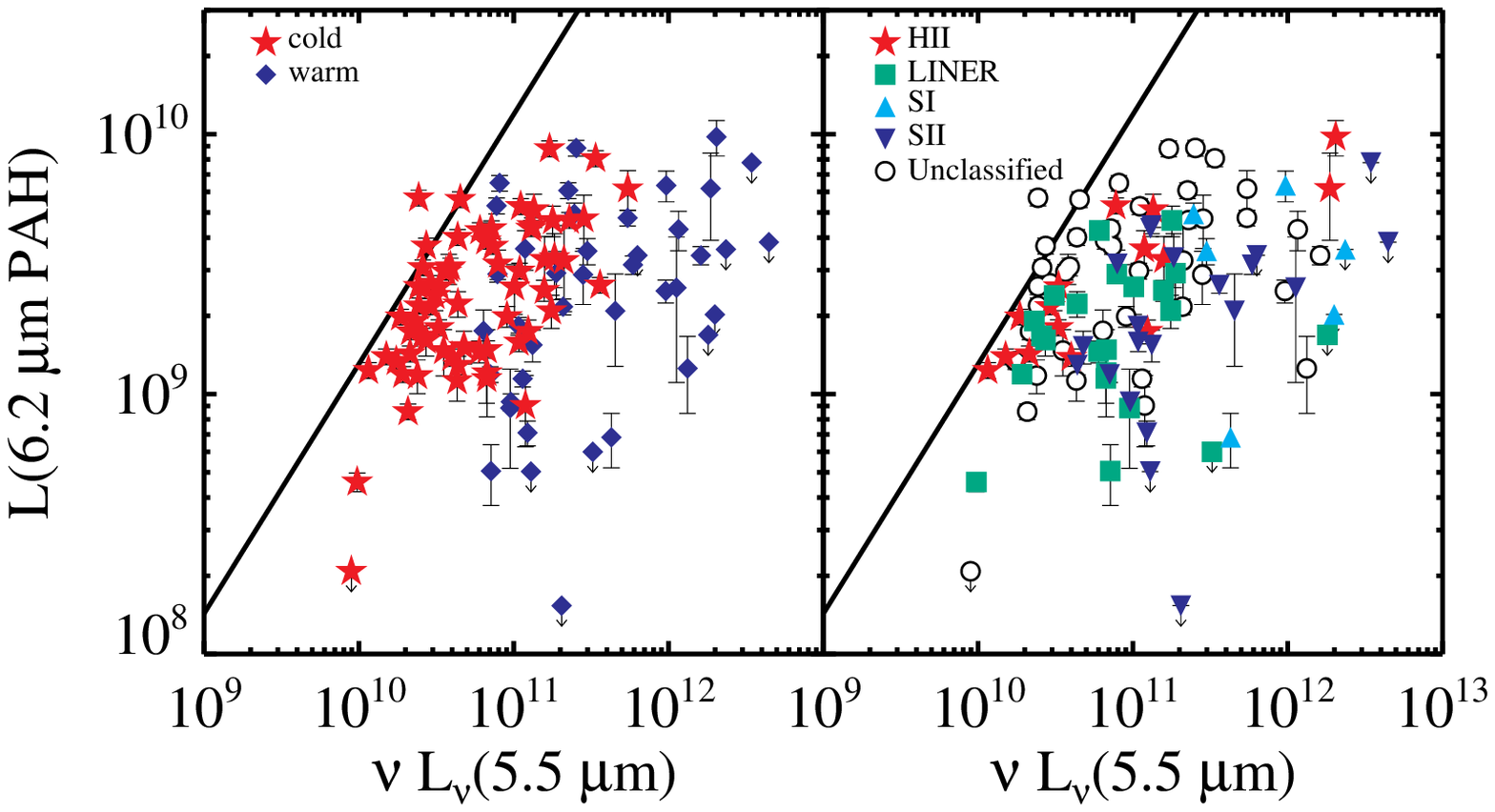}}
\bigskip
\bigskip
\centerline{
\includegraphics[scale=1.0]{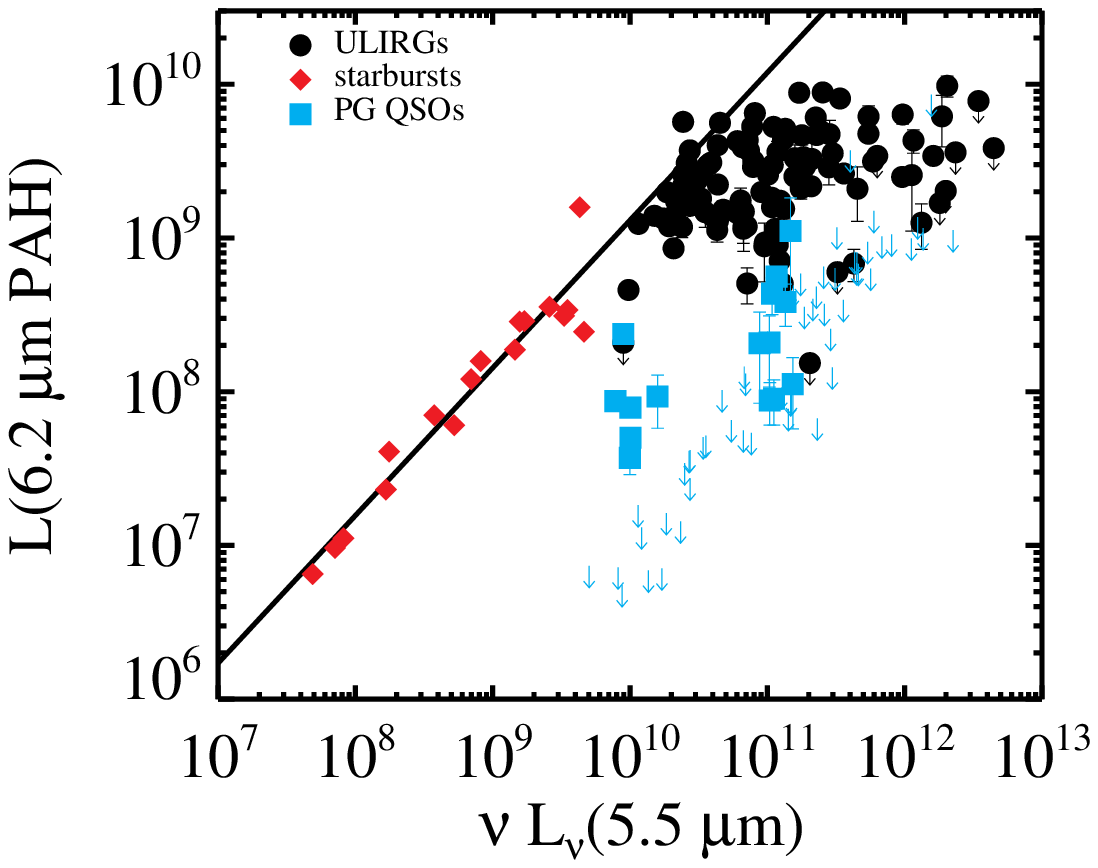}}

\caption{Luminosity in the 6.2 $\micron$ PAH feature as a function of
  5.5 $\micron$ luminosity.  In the top panels, the ULIRGs are
  color-coded according to spectral slope (left) and optical
  classification (right) using the same symbols as in Figure
  \ref{fig:pahew_v_lum24}.  The solid black line is a fit to the
  starbursts analyzed by \citet{Brandl06}.  The bottom panel features
  expanded x and y axes to show the locations of the \citet{Brandl06}
  starbursts (red diamonds) and $\sim$80 $z < 0.5$ PG QSOs, displayed
  as cyan squares (detections) and arrows (upper limits).  In this
  bottom panel, the ULIRGs are shown as black circles.}

\label{fig:pahlum_v_lum5p5}
\end{figure}


\begin{figure}
\centerline{
\includegraphics[scale=1.0]{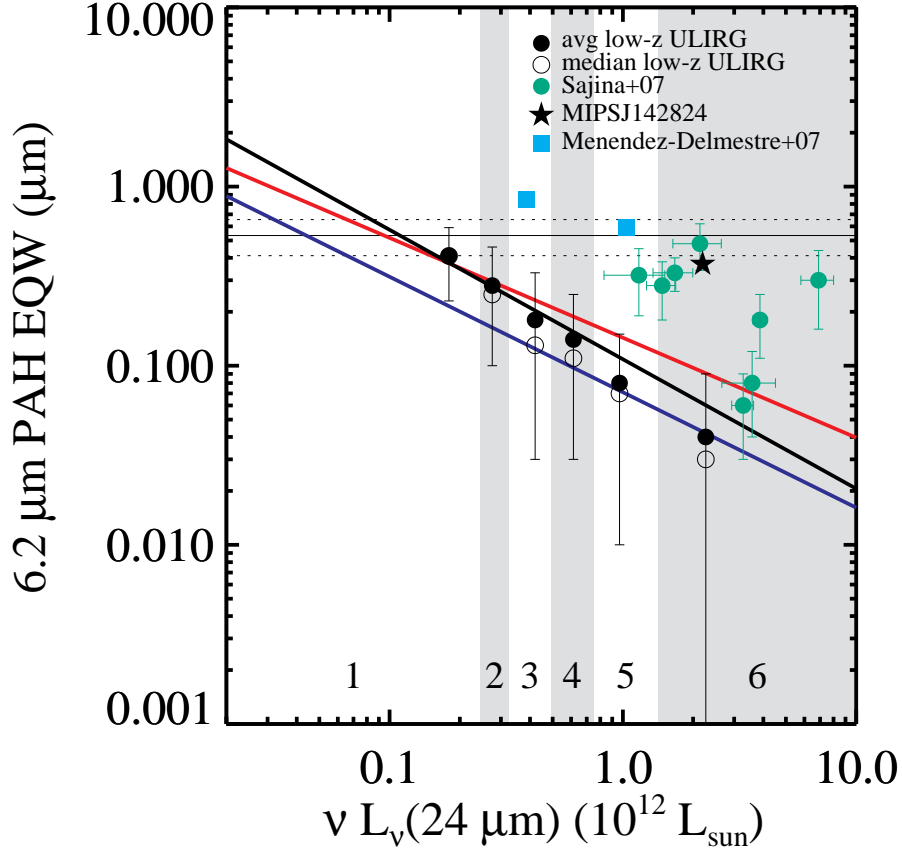}}

\caption{Median (open circles) and average (filled circles) 6.2
$\micron$ PAH EQW versus 24 $\micron$ luminosity.  Both the medians
and the means were computed by including sources with PAH EQW upper
limits in the same way as detections.  The error bars were computed as
the dispersion in the PAH EQWs in each luminosity bin, and are
centered on the average values in this plot.  The x-position of each
point is the median of the 24 $\micron$ luminosities of the sources in
each luminosity bin.  The thick red, blue, and black lines are fits to
the cold ULIRGs, the warm ULIRGs, and all of the ULIRGs, respectively.
Sources with upper limits on the PAH EQW were included in the fits as
detections with 10\% error bars.  The horizontal thin and dotted black
lines have the same meaning as in Figure \ref{fig:pahew_v_lum24}, as
do the luminosity bin labels across the bottom.  The green points are
the high-redshift ($z \sim 2$) ULIRGs from \citet{Sajina07} with
measured 6.2 $\micron$ PAH EQWs, and rest-frame 24 $\micron$
luminosities estimated from the 70 $\micron$ MIPS photometry.  The
large black star is MIPS J142824.0+352619, a bright 160 $\micron$
source at $z=1.3$.  The rest-frame 24 $\micron$ luminosity for this
object was computed by via log-log interpolation of the photometry
presented in \citet{Borys06}, and the 6.2 $\micron$ PAH EQW was taken
from \citet{Desai06}.}

\label{fig:pahew_v_lum24_medians}
\end{figure}

\end{document}